\documentclass[fleqn,usenatbib]{mnras}

\usepackage{newtxtext,newtxmath}

\usepackage[T1]{fontenc}
\usepackage{tabularx}
\usepackage{booktabs}
\allowdisplaybreaks

\DeclareRobustCommand{\VAN}[3]{#2}
\let\VANthebibliography\thebibliography
\def\thebibliography{\DeclareRobustCommand{\VAN}[3]{##3}\VANthebibliography}

\usepackage{graphicx}	

\def\m87{{M87$^{\ast} $}}

\newcommand{\adi}{\Gamma}

\newcommand{\mypar}{{\mkern3mu\vphantom{\perp}\vrule depth 0pt\mkern3mu\vrule depth 0pt\mkern3mu}}
\newcommand{\vpar}{\tilde{v}_\mypar}
\newcommand{\vperp}{\tilde{v}_\perp}
\newcommand{\upar}{\tilde{u}_\mypar}
\newcommand{\uperp}{\tilde{u}_\perp}
\newcommand{\ugas}{u_\text{gas}}
\newcommand{\lpar}{\gamma_\mypar}
\newcommand{\lperp}{\gamma_\perp}

\newcommand{\B}{\mathcal{B}}
\newcommand{\E}{\mathcal{E}}
\newcommand{\Fd}{\vphantom{F}^\star \! F}

\def\lsim{\mathrel{\raise.3ex\hbox{$<$\kern-.75em\lower1ex\hbox{$\sim$}}}}
\def\gsim{\mathrel{\raise.3ex\hbox{$>$\kern-.75em\lower1ex\hbox{$\sim$}}}}
\def\gtwid{\mathrel{\raise.3ex\hbox{$>$\kern-.75em\lower1ex\hbox{$\sim$}}}}
\def\proptwid{\mathrel{\raise.3ex\hbox{$\propto$\kern-.75em\lower1ex\hbox{$\sim$}}}}

\graphicspath{{./}{figures/}}

\title[Hybrid GRMHD+GRFFE Simulations]{Hybrid GRMHD and Force-Free Simulations of Black Hole Accretion}

\author[A. Chael]{
Andrew Chael$^{1}$\thanks{E-mail: achael@princeton.edu}
\\
$^{1}$Princeton Gravity Initiative, Princeton University, Princeton NJ, 08540\\
}

\date{Accepted XXX. Received YYY; in original form ZZZ}

\pubyear{2024}

\begin{document}
\label{firstpage}
\pagerange{\pageref{firstpage}--\pageref{lastpage}}
\maketitle

\begin{abstract}
We present a new approach for stably evolving general relativistic magnetohydrodynamic (GRMHD) simulations in regions where the magnetization $\sigma=b^2/\rho c^2$ becomes large. GRMHD codes typically struggle to evolve plasma above $\sigma\approx100$ in simulations of black hole accretion. To ensure stability, GRMHD codes will inject mass density artificially to the simulation as necessary to keep the magnetization below a ceiling value $\sigma_{\rm max}$. We propose an alternative approach where the simulation transitions to solving the equations of general relativistic force-free electrodynamics (GRFFE) above a magnetization $\sigma_{\rm trans}$.  We augment the GRFFE equations in the highly magnetized region with approximate equations to evolve the decoupled field-parallel velocity and plasma energy density. Our hybrid scheme is explicit and easily added to the framework of standard-volume GRMHD codes. We present a variety of tests of our method, implemented in the GRMHD code \texttt{KORAL}, and we show results from a 3D hybrid GRMHD+GRFFE simulation of a magnetically arrested disc (MAD) around a spinning black hole. Our hybrid MAD simulation closely matches the average properties of a standard GRMHD MAD simulation with the same initial conditions in low magnetization regions, but it achieves a magnetization $\sigma\approx10^6$ in the evacuated jet funnel. We present simulated horizon-scale  images of both simulations at 230 GHz with the black hole mass and accretion rate matched to M87*. Images from the hybrid simulation are less affected by the choice of magnetization cutoff $\sigma_{\rm cut}$ imposed in radiative transfer than images from the standard GRMHD simulation.
\end{abstract}




\section{Introduction}

General Relativistic Magnetohydrodynamic (GRMHD) codes are now a standard tool for investigating accretion flows, jets, and outflows on black hole horizon scales. 
Most supermassive black holes accrete slowly, with an accretion rate $\dot{M}$ many orders of magnitude below the Eddington limit $\dot{M}_{\rm Edd}$. The accreting plasma around these black holes is hot ($T>10^{10} K$), dilute, and optically thin \citep{Yuan14}. These flows can be strongly magnetized and frequently launch jets by extracting the spin energy of the black hole \citep{BZ}. 

Millimeter-wave synchrotron emission from hot accretion flows around supermassive black holes is produced on horizon scales at radii $r < 10 \, r_{\rm g}$, where $r_{\rm g}=GM/c^2$ is the black hole's gravitational radius. Using Very Long Baseline Interferometry (VLBI) at 230 GHz, the Event Horizon Telescope (EHT) has resolved and imaged this horizon-scale emission around the supermassive black holes with the largest apparent sizes: M87* \citep{PaperIV} and Sgr A* \citep{SgrAPaperIII}. 
The EHT images of both sources have been extensively compared to simulated images from GRMHD simulation models made using polarized, relativistic ray-tracing radiative transfer codes \citep[e.g.][]{Dexter16,ipole}. 
Connecting EHT images to simulated data from GRMHD simulations has indicated that the emitting plasma around M87* co-rotates with the black hole \citep{PaperV} and that it is likely in the magnetically arrested (MAD) state of black hole accretion \citep{PaperVIII, PaperIX}. EHT observations and simulations of Sgr A* also indicate that it is likely magnetically arrested \citep{SgrAPaperV,SgrAPaperVII,SgrAPaperVIII}. 

The strongly magnetized regime is particularly interesting for understanding the physics of relativistic jet launching, plasma heating, and flaring processes around supermassive black holes. Unfortunately, numerical 
GRMHD codes struggle when the magnetic field energy density greatly exceeds the plasma rest mass density, or when the magnetization parameter $\sigma\gg1$:
\begin{equation}
\label{eq:sigma}
\sigma\equiv\frac{b^2}{\rho c^2}.
\end{equation}
Here, $b^2$ is the magnetic energy density in the fluid rest frame in Heaviside-Lorentz units,\footnote{Heaviside-Lorentz units are related to Gaussian units by a factor of $4\pi$ in the magnetic energy density, such that $b^2_{\rm HL}=b^2_{\rm G}/4\pi$.} and $\rho$ is the fluid mass density. 
In highly magnetized regions, the fluid rest mass and thermal energy make up a small contribution to the overall energy budget. Small numerical errors in the evolution of the overall energy-momentum can cascade to large errors in the fluid quantities, and the simulation can evolve to an unphysical state and crash. 

The typical solution to the instability of GRMHD codes at high $\sigma$ is to introduce numerical ``floors'' on the mass density $\rho$. In practice, this is usually achieved by limiting $\sigma<\sigma_{\max}\approx100$ throughout the simulation by constantly increasing the density in regions that would otherwise prefer to climb to higher magnetization. The details of this flooring procedure, such as which frame mass is added to, can affect the results of the simulation \citep{Ressler17}.
Furthermore, when producing simulated synchrotron images from GRMHD simulations for comparison to observations, it is necessary to remove emission from ``floored'' regions with an artificially high plasma density, or $\sigma>\sigma_{\rm max}$. In practice, most studies of GRMHD images more conservatively remove all emission from regions with $\sigma>\sigma_{\rm cut}\approx 1$, as the evolution of the plasma internal energy and temperature is considered to be untrustworthy even below the ceiling value $\sigma_{\rm max}$. The resulting millimeter-wavelength image structure can be sensitive to the chosen value of $\sigma_{\rm cut}$ \citep[e.g.][]{Chael19,Zhang2024}.

This paper proposes an alternative approach to density floors in GRMHD simulations. Instead of artificially increasing the density to force the simulation to stay below a maximum magnetization $\sigma_{\max}$, we present a method to smoothly connect the evolution of the GRMHD equations to their force-free limit in regions of high $\sigma$.
The equations of general relativistic force-free electrodynamics (GRFFE) represent the dynamics of an infinitely conductive, zero-rest-mass magnetized plasma in curved spacetime and are the limit of the GRMHD equations as $\sigma\rightarrow\infty$. The GRFFE equations are typically solved by evolving the magnetic and electric fields in some coordinate system \citep[e.g.][]{Komissarov02b,McKinney06,Etienne2017,Mahlmann2021}. However,
\citet{McKinney06} showed that the GRFFE equations can be solved in standard finite-volume GRMHD codes with only small modifications, essentially removing the back-reaction of the fluid variables on the evolution of the magnetic field and plasma velocity. 

In our approach, we adopt the method of \citet{McKinney06} to solve the equations of GRFFE in highly magnetized regions above a transition magnetization $\sigma>\sigma_{\rm trans}$. Since the GRFFE equations do not constrain the evolution of the fluid density and temperature, nor the fluid velocity parallel to the magnetic field lines, we solve additional equations to approximately evolve these quantities in the high-$\sigma$ or GRFFE region. We then smoothly connect the GRFFE evolution region in the magnetized jet region of a black hole accretion simulation to standard GRMHD evolution in most of the simulation volume. 

Other works have considered different methods for connecting GRMHD and GRFFE evolution in studies of neutron star magnetospheres. \citet{Paschalidis13} switch between GRMHD and GRFFE evolution at a fixed boundary, the neutron star surface; as they are interested primarily in the evolution of the neutron star magnetosphere, they do not attempt to evolve fluid quantities in this region. \citet{Parfrey2017} take a different approach, adapting a GRMHD code as in this work to operate in the high $\sigma$ limit by manually fixing the fluid density $\rho$ and internal energy $\ugas$ to pre-determined values in the magnetosphere and solving for the evolution of the electromagnetic field by the GRMHD equations on this fixed background. In our approach, we solve directly for the magnetic field, velocity, and plasma density and internal energy in both regions, under different equations. Like \citet{Parfrey2017}, we allow the boundary between the GRMHD and GRFFE regions to evolve dynamically throughout the simulation. We find that our approach can be straightforwardly applied to GRMHD simulations of MAD accretion discs, and that in these simulations we can stably evolve fluid with magnetizations $\sigma>10^6$ in the jet region close to the black hole. Our approach affects near-horizon images obtained from GRMHD simulations less than the standard floor approach, as the density in the uncertain high-$\sigma$ regions naturally falls to zero. 

The paper is organized as follows. First, we review the GRMHD and GRFFE equations and their properties in \autoref{sec:equations}. Then in \autoref{sec:method} we review the standard finite-volume method for evolving the GRMHD equations, its extension to force-free evolution, and our new approach for coupling the two in different regions of a simulation. In \autoref{sec:tests} we present several tests of our method, implemented in the GRMHD code \texttt{KORAL} \citep{KORAL13}. In \autoref{sec:simulations} we present a comparison between two 3D MAD simulations performed with a standard GRMHD method including density floors and our new hybrid scheme. We discuss and conclude in \autoref{sec:discussion}. 

\section{GRMHD and GRFFE Equations}
\label{sec:equations}
\subsection{Units and Definitions}

Throughout, we use units normalized such that $c=G=1$. We work in a fixed background metric $g_{\mu\nu}$ with signature $(-,+,+,+)$, where we take the zeroth coordinate to be timelike at spatial infinity and we take Latin indices to refer to spatial components. 
Throughout, we use the ADM form of the metric: 
\begin{equation}
    ds^2 = -\alpha^2 \mathrm{d}x^0\mathrm{d}x^0 + \sigma_{ij}(\mathrm{d}x^i + \beta^i \mathrm{d}x^0)(\mathrm{d}x^j + \beta^j \mathrm{d}x^0),
\end{equation}
where the lapse $\alpha$, shift vector $\beta^i$, and spatial metric $\sigma_{ij}$ are
\begin{align}
    \alpha^2 &= -1/g^{00} \;\;,\;\; \beta^i = \alpha^2g^{0i }\;\;,\;\; \sigma_{ij} = g_{ij}.
\end{align}

We frequently work in the normal observer frame $\eta^\mu$ for a given coordinate system. The normal observer frame is defined by the covariant timelike vector $\eta_\mu = \left(-\alpha,0,0,0\right)$, so that
\begin{equation}
    \eta^\mu = \left(1/\alpha,-\beta^i/\alpha\right),
\end{equation}
We define a projection tensor $j^{\mu}_{\;\nu}$ into the normal observer frame:
\begin{equation}
j^\mu_{\nu} = \delta^\mu_{\;\nu} + \eta^\mu \eta_\nu. 
\end{equation}
Given a four-velocity $u^\mu$, we can then find the projected four-velocity in the normal observer frame $\tilde{u}^\mu = j^\mu_{\;\nu} u^\nu$. The projected normal-observer four-velocity has components
\begin{align}
    \tilde{u}^0 &= 0 \;\;,\;\; \tilde{u}^i = u^i - \gamma\eta^i,
    \label{eq:normalobsvel}
\end{align}
where the Lorentz factor $\gamma$ defined in the normal observer frame is
\begin{equation}
    \gamma = -\eta_\mu u^\mu = \alpha u^0.
\end{equation}
We can also compute the Lorentz factor from the normal observer frame three-velocity $\tilde{v}^i=\tilde{u}^i/\gamma$ by the familiar expression
\begin{equation}
    \gamma = \frac{1}{\sqrt{1-\tilde{v}^2}},
    \label{eq:gammafromv}
\end{equation}
where $\tilde{v}^2 = \sigma_{ij}\tilde{v}^i\tilde{v}^j$.\footnote{Furthermore, $\gamma=\sqrt{1+\tilde{u}^2}$, where $\tilde{u}^2 = \sigma_{ij}\tilde{u}^i\tilde{u}^j$. Note that to \emph{lower} indices on contravariant normal observer frame 3-vectors we can use the spatial metric $\sigma_{ij}=g_{ij}$, but to \emph{raise} indices on covariant normal frame 3-vectors we must use $\sigma^{ij} = g^{ij} + \frac{1}{\alpha^2}\beta^i\beta^j.$} The projection of the four-velocity into the normal observer frame, \autoref{eq:normalobsvel}, is invertible, so that given $\tilde{u}^i$ or $\tilde{v}^i$ we can solve for $\gamma$ using \autoref{eq:gammafromv}, and then determine the four-velocity from $u^\mu=\tilde{u}^\mu+\gamma\eta^\mu$. 

An electromagnetic field is defined by its antisymmetric field strength (Faraday) tensor $F^{\mu\nu}$, and its dual (Maxwell) tensor  $\Fd^{\mu\nu}$.\footnote{The Hodge dual in four dimensions is $\Fd^{\mu\nu} = \frac{1}{2}\epsilon^{\mu\nu\kappa\lambda}F_{\kappa\lambda}$. The Levi-Civita tensor $\epsilon^{\alpha \beta\gamma\delta} = (-1/\sqrt{-g})[\alpha\beta\gamma\delta]$, while $\epsilon_{\alpha \beta\gamma\delta} = \sqrt{-g}[\alpha\beta\gamma\delta]$, where $[\alpha\beta\gamma\delta]$ is the completely antisymmetric symbol.} 
We decompose the field tensors into coordinate-dependent electric and magnetic field four-vectors in the normal observer frame:
\begin{align}
\E^\mu &= \eta_\nu F^{\mu\nu}, \\
\B^\mu &=-\eta_\nu \Fd^{\mu\nu}.
\label{eq:normalobsfields}
\end{align}
By definition, the contravariant normal observer electric and magnetic field vectors\footnote{GRMHD codes most often work with the ``lab frame'' electric and magnetic field three-vectors $E^i=F^{0i}=\E^i/\alpha$ and $B^i=-\Fd^{0i}=\B^i/\alpha$ instead of the normal observer frame fields \citep[e.g.][]{Gammie03}.}
have no time component, $\E^0=\B^0=0$, and they are both orthogonal to $\eta^\mu$.

\subsection{GRMHD Equations}

Here, we review the equations of ideal GRMHD following \citet{Gammie03} and \citet{McKinney06}. Ideal magnetohydrodynamics describes a single perfect fluid coupled to a degenerate electromagnetic field. The perfect fluid stress energy tensor is
\begin{equation}
    T^{\mu\nu}_{\mathrm{fluid}} = h\, u^\mu u^\nu + p g^{\mu\nu},
    \label{eq:Tfluid}
\end{equation}
where $u^\mu$ is the timelike four-velocity,  $h$ is the relativistic enthalpy density, and $p$ is the pressure, both measured in the fluid rest frame.  
We adopt an adiabatic equation of state with index $\adi$ such that $p=(\adi-1)\ugas$, where $\ugas$ is the fluid internal energy. The enthalpy density $h$ is then
\begin{equation}
    h = \rho + \ugas + p = \rho + \Gamma \ugas,
\end{equation}
where $\rho$ is again the fluid rest-mass density. The dimensionless temperature is 
\begin{equation}
    \Theta_{\rm gas}=\frac{p}{\rho}.
\end{equation}

To take the ideal MHD approximation, we assume that the fluid's conductivity is infinite and hence the electric field is zero in the fluid rest frame, $e^\mu = u_\nu F^{\mu\nu}=0$. The magnetic field in the fluid rest frame is generally nonzero and is given by
\begin{equation}
    b^\mu = -u_\nu \Fd^{\mu\nu},
\end{equation}
which is orthogonal to $u^\mu$ by the antisymmetry of $\Fd^{\mu\nu}$.
The fluid-frame magnetic field vector $b^\mu$ is related to the normal observer frame magnetic field $\B^\mu$ by a projection:\footnote{Since the field is degenerate, the normal observer frame electric and magnetic fields are perpendicular $\E^\alpha \B_\alpha=0$.} 
\begin{equation}
 b^\mu = \frac{1}{\gamma}\left(\delta^{\mu}_{\;\nu}+u^\mu u_\nu\right)\B^\nu. 
 \label{eq:bmu}
\end{equation}

The Faraday tensor in GRMHD has a nontrivial kernel; that is, there exists at least one frame $u^\mu$ where $u_\nu F^{\mu\nu}=0$, so that the electric field in that frame vanishes. In general, when $F^{\mu\nu}$ has a nontrivial kernel, the electromagnetic field is called \emph{degenerate}. When at least some of the frames $u^\mu$ in the kernel are timelike, the field is \emph{magnetically dominated}.

When the electromagnetic field is both degenerate and magnetically dominated, the Maxwell tensor can be expressed simply in terms of the velocity $u^\mu$ and fluid frame magnetic field $b^\mu$ as
\begin{equation}
    \Fd^{\mu\nu} = b^\mu u^\nu - u^\mu b^\nu. 
    \label{eq:MaxDeg}
\end{equation}
The stress-energy tensor for a degenerate, magnetically dominated electromagnetic field is
\begin{equation}
    T^{\mu\nu}_\mathrm{EM} = b^2u^\mu u^\nu + \frac{1}{2}b^2g^{\mu\nu} - b^\mu b^\nu.
    \label{eq:TemDeg}
\end{equation}
In the Appendix, \autoref{app:fields}, we review the forms of $F^{\mu\nu}$, $\Fd^{\mu\nu}$, and $T_{\rm EM}^{\mu\nu}$ in terms of the normal observer frame fields.

Given these definitions, the GRMHD equations of motion can be expressed simply as the conservation of the total stress-energy $T^{\mu\nu}=T^{\mu\nu}_\mathrm{fluid} + T^{\mu\nu}_\mathrm{EM}$:
\begin{equation}
  \nabla_\mu T^{\mu\nu} = 0,
  \label{eq:GRMHDT}
\end{equation}
along with the conservation of the mass density current
\begin{equation}
  \nabla_\mu \left(\rho u^\mu\right) = 0,
    \label{eq:advection}
\end{equation}
and the homogeneous Maxwell equation
\begin{equation}
   \nabla_\mu \Fd^{\mu\nu} = 0. 
     \label{eq:MaxHom}
\end{equation}
The inhomogeneous Maxwell equation $\nabla_\mu F^{\mu\nu}= -J^\nu$ is not required to evolve the GRMHD system of equations, and it is instead taken to define the electric current $J^\nu$.  

\subsection{Force-Free Equations}

When the electromagnetic stress-energy dominates over the fluid stress-energy, $T^{\mu\nu}_{\rm EM} \gg T^{\mu\nu}_{\rm fluid}$, the GRMHD equations for the electromagnetic field become independent of the fluid: 
\begin{align}
\label{eq:FFE1}
  \nabla_\mu T^{\mu\nu}_\mathrm{EM} &= 0, \\  
  \nabla_\mu\Fd^{\mu\nu} &= 0. 
  \label{eq:FFE}
\end{align}
These are the equations of force-free electrodynamics (GRFFE). 

In both GRFFE and ideal GRMHD, the electromagnetic field is degenerate and magnetically dominated.
One can show \citep[e.g.][]{McKinney06} that if there is at least one timelike frame $u^\mu$ with vanishing electric field, $e^\mu=u_\nu F^{\mu\nu}=0$, the electric field also vanishes in an infinite family of timelike frames. These frames are connected by Lorentz boosts along magnetic field lines, and \autoref{eq:MaxDeg} and \autoref{eq:TemDeg} are valid descriptions of the Maxwell and stress-energy tensors in any of these frames.\footnote{In this section we use $b^\mu$ to refer to the magnetic field in any timelike frame $u^\mu$; beginning in \autoref{sec:method}, $b^\mu$, we reserve $b^\mu$ to refer to the magnetic field in the fluid rest frame.}
In GRMHD, the fluid velocity picks out a unique $u^\mu$ with vanishing electric field, but in GRFFE the evolution equations alone do not pick out a unique $u^\mu$. 

Among the infinite number of frames with vanishing electric field, the \emph{drift frame} $u^\mu_\perp$ is the unique frame with zero electric field where the Lorentz factor relative to the normal observer $\eta^\mu$ is minimized:
\begin{equation}
    u^\mu_\perp = \gamma_\perp \left[\eta^\mu - \frac{\epsilon^{\mu\nu\alpha\beta}\eta_\nu \E_\alpha \B_\beta}{\B^2}\right].
    \label{eq::uperp}
\end{equation}
The Lorentz factor of the drift frame is 
\begin{equation}
 \gamma_\perp = \sqrt{\frac{\B^2}{\B^2-\E^2}}. 
 \label{eq:lperp}
\end{equation}

It is clear from the form of \autoref{eq:lperp} that the field must be magnetically dominated for the drift velocity in the normal observer frame to be slower than the speed of light. Since all other frames with vanishing electric field have a larger Lorentz factor than the drift frame, the field must be magnetically dominated for any frame with vanishing $e^\mu$ to be timelike.

Given a magnetically dominated degenerate field and a fixed coordinate system, the drift frame is the unique frame where the electric field vanishes, where the Lorentz factor relative to the normal observer is minimized, and which is orthogonal to the normal-observer magnetic field, $u_{\perp\nu} \B^\nu = \eta_\mu u_{\nu\perp}\Fd^{\mu\nu}=0$. Because the drift frame depends on the choice of normal-observer frame $\eta^\mu$, it is not a ``physical'' frame. Unlike the fluid frame in GRMHD, the drift frame and the drift frame Lorentz factor change depending on the choice of coordinate system.

All other frames with vanishing electric field can be parameterized by their three-velocity $\vpar$  along the magnetic field line in the normal observer frame, which is encoded in the zeroth component of the magnetic field $b^\mu$:
\begin{equation}
    b^0 = \frac{1}{\alpha}u_\mu \B^\mu = \frac{\gamma}{\alpha}\vpar\sqrt{\B^2}. 
\label{eq:b0}
\end{equation}
A general frame where the electric field $e^\mu$ vanishes is thus related to the drift frame $u^\mu_\perp$ by a Lorentz boost along the normal observer magnetic field:
\begin{equation}
 u^\mu = \gamma\left(\frac{1}{\gamma_\perp}u^\mu_\perp + \vpar \frac{\B^\mu}{\sqrt{\B^2}}\right).
 \label{eq:ugeneral}
\end{equation}
To ensure that \autoref{eq:ugeneral} remains timelike, the parallel three-velocity has a maximum value
\begin{equation}
\vpar^2 < 1-\E^2/\B^2.
\end{equation}
Alternatively, we can parameterize the boost in terms of a parallel Lorentz factor $\lpar\geq 1$  such that the total Lorentz factor is 
\begin{equation}
    \gamma=\left(\lperp^{-2}+ \lpar^{-2}-1\right)^{-1/2}.
\end{equation}
In the Appendix, \autoref{app:velpar}, we present a more detailed discussion of the field-parallel and field-perpendicular velocities. 

The drift frame and the parallel boost Lorentz factor are coordinate-dependent quantities defined in terms of the normal observer $\eta_\mu$. In different coordinate systems for the same metric (e.g. Boyer-Lindquist and Kerr-Schild coordinates for the Kerr metric of a spinning black hole), identical degenerate EM fields will have different drift velocities, depending on $\eta^\mu$. 
When solving the force-free equations \ref{eq:FFE1} and \ref{eq:FFE} numerically in terms of a velocity $u^\mu$ and magnetic field $b^\mu$ , the drift frame velocity is typically used, since the GRFFE equations do not determine the velocity parallel to the magnetic field lines \citep{McKinney06}.

\section{Coupling GRMHD and GRFFE in finite volume codes }
\label{sec:method}

\subsection{Finite Volume GRMHD Evolution}

Finite-volume GRMHD codes define the fluid and magnetic field in terms of a vector of ``primitive'' variables $\mathbf{P}$ defined in each cell of a discretized grid. The eight primitive quantities in GRMHD are
\begin{equation}
    \mathbf{P} = \left[\rho,\; \ugas, \; \tilde{u}^i, \;\B^i \right].
    \label{eq:prims}
\end{equation}
GRMHD codes usually use the normal observer frame velocity components $\tilde{u}^i$ as a primitive quantity instead of $u^i$, since they can take any value from $-\infty$ to $\infty$ and still produce a timelike $u^\mu$ \citep{McKinney2004}.

The GRMHD equations in \autoref{eq:GRMHDT}--\autoref{eq:MaxHom} can then all be expressed in finite volume form
\begin{equation}
\partial_0 \mathbf{U}(\mathbf{P}) = -\partial_j \mathbf{F}^j(\mathbf{P}) + \mathbf{S}(\mathbf{P}),
\label{eq:fluxes}
\end{equation}
where $\mathbf{U}$ are the ``conserved'' quantities, $\mathbf{F}$ are the fluxes between spatial cells, and $\mathbf{S}$ are source terms. The conserved quantities in GRMHD are 
\begin{align}
    \mathbf{U} 
     &= \frac{\sqrt{-g}}{\alpha} \times \left[\mathcal{D},\; \mathcal{Q}_0,\;  \mathcal{Q}_i, \; \B^i \right],
     \label{eq:conserved}
\end{align}
where $\mathcal{D}=\gamma \rho$ is the density in the normal observer frame and $\mathcal{Q}_\mu = -\eta_\nu T^{\nu}_{\;\;\mu} = \alpha T^{0}_{\;\;\mu}$ is the energy flux four-vector in the normal observer frame. The fluxes are
\begin{align}
    \mathbf{F}^j &= \sqrt{-g}\times\left[\rho u^j, T^j_{\;0}, T^j_{\;i}, \left(b^ju^i-b^iu^j\right)\right].
    \label{eq:fluxes2}
\end{align}
In ideal GRMHD without additional physics (e.g. coupling to radiation), the source terms $\mathbf{S}$ arise purely from the geometry; they are nonzero only for the stress-energy equations and involve products of the Christoffel symbols \citep[see][Eq 4]{Gammie03}. 

The GRMHD equations are hyperbolic and for a given coordinate system they may be evolved forward in the time coordinate $t=x^0$ explicitly. A timestep $\Delta t$ begins by interpolating the primitive quantities $\mathbf{P}$ to cell walls. The conserved quantities and source terms in each cell and the flux terms on each cell wall are then computed. The conserved quantities are updated by summing up the fluxes entering and leaving the cell, along with the geometric source terms, multiplied times $\Delta t$.    

After the conserved quantities $\mathbf{U}$ are explicitly evolved forward in time with \autoref{eq:fluxes}, a GRMHD code must solve for the primitives $\mathbf{P}(\mathbf{U})$ in each cell.
In GRMHD, the map from the conserved quantities $\mathbf{U}$ to the primitives $\mathbf{P}$ is not analytically invertable, so the conservative-to-primitive inversion must be done numerically. 
\citet{Noble2006} showed that the primitives $\mathbf{P}$ can all be expressed analytically in terms of the conserved quantities and the relativistic enthalpy $W\equiv\gamma^2 h$. Most GRMHD codes numerically solve for this single variable, $W$, from the conserved quantities using a Newton-Raphson approach. We review the \citet{Noble2006} inversion procedure in the Appendix, \autoref{app:u2p} 

Unfortunately, GRMHD numerical inversion can fail.
When the magnetization $\sigma$ is large, the magnetic parts of the conserved quantities $\tilde{\mathcal{Q}}^i$ and $\mathcal{U}\equiv-\eta_\mu\mathcal{Q}^\mu$ dominate the contributions from the fluid energy momentum (\autoref{eq:QandUv2}). As a result of truncation error, a GRMHD code can end up in a situation where no consistent solution for $W$ can be found given the numerically evolved conserved quantities. 
GRMHD codes handle these failures by imposing artificial ceilings on $\sigma$, or equivalently floors on the density $\rho$ These floors are handled differently in different codes, but they all have the effect of limiting the magnetization below some ceiling $\sigma < \sigma_{\rm max}$ in some frame. Except for some very well-behaved problems, $\sigma_{\rm max} \approx 100$ in most applications in most GRMHD codes.

The region of parameter space where GRMHD conserved-to-primitive inversion fails, as $\sigma$ becomes large, is also where the GRMHD equations approach their force-free limit. Thus, instead of imposing floors on $\rho$, we may hope to handle the failure of GRMHD codes at large $\sigma$ by transitioning to solving the GRFFE equations in regions of large $\sigma$. However, since the GRFFE equations only uniquely determine the drift frame velocity $u_\perp^\mu$ and its associated magnetic field $b^\mu_\perp$, we will have to add additional equations to determine the evolution of the density, fluid internal energy, and parallel velocity.

\subsection{Finite Volume GRFFE Evolution}
\label{sec:grffe}

Most force-free codes determine the evolution of the electromagnetic field $F^{\mu\nu}$ under the GRFFE equations \autoref{eq:FFE} using the electric and magnetic field components $\{\B^i,\E^i \}$ as primitive variables. \citet{McKinney06} showed that it is possible to evolve the force-free equations of motion in the same framework as a standard GRMHD code, where the primitive variables are the magnetic field components $B^i$ and drift frame velocity $\tilde{u}_\perp^i$.

In this approach to GRFFE, the primitive variables are $\mathbf{P}_{\rm FFE} = \left[\tilde{u}^i_\perp, \B^i\right]$. The associated conserved quantities, fluxes, and source terms are the same as in GRMHD, except that only $T^{\mu\nu}_{\rm EM}$ is used in \autoref{eq:conserved} and \autoref{eq:fluxes2} instead of the combined $T^{\mu\nu}_{\rm EM}+T^{\mu\nu}_{\rm fluid}$.

Unlike in GRMHD, the conserved-to-primitive inversion in this formulation of GRFFE is analytic. Specifically, \citet{McKinney06} showed that from the conserved quantities $\B^i$ and $\mathcal{Q}_i = \alpha T^0_{\;i}$ obtained from finite volume evolution, the normal observer electric field is
\begin{equation}
  \E^\alpha = -\frac{\epsilon^{\alpha\beta\gamma\delta}\, \B_\beta\, \mathcal{Q}_\gamma \, \eta_\delta}{\B^2},
  \label{eq:EMcKinney}
\end{equation}
and then the drift-frame velocity $u_\perp^\mu$ can be obtained from \autoref{eq::uperp}. 
In fact, the \citet{McKinney06} solution \autoref{eq:uMcKinney} is equal to what is obtained by setting $W=0$ in the usual \citet{Noble2006} GRMHD inversion procedure (\autoref{eq:Qofv}). Namely, given the energy flux projected in the normal observer frame $\tilde{Q}^i$, the drift three-velocity is
\begin{equation}
    \vperp^i = \frac{\tilde{Q}^i}{\B^2},
    \label{eq:vperpFF}
\end{equation}
and the normal observer velocity is then $\uperp^i=\lperp\vperp^i$, where $\lperp$ is calculated with \autoref{eq:gammafromv}.

Force-free electrodynamics can be implemented in this adapted GRMHD finite volume framework as long as the field remains magnetically dominated, $\B^2 > \E^2$. In practice, a set of GRFFE initial conditions are not guaranteed to remain magnetically dominated as they evolve. In particular, in current sheets the field can become electrically dominated as the assumptions of GRFFE break down. We thus need to implement a ceiling on $\lperp<\gamma_{\rm max}$ to keep the field magnetically dominated and the evolution stable throughout the simulation region. Such a ceiling on $\gamma$ is also standard in GRMHD codes.

\subsection{Decoupled GRFFE \& Fluid Evolution}
\label{sec:decouple}

Next, we wish to define a strategy for fully evolving a fluid and electromagnetic field in the force-free limit, where the fluid does not back-react on the evolution of the field. Given a GRFFE solution for $\B^\mu$ and $u^\mu_\perp$, we can couple a non-interacting fluid to the force-free field by adding three evolution equations for the fluid mass density $\rho$, the fluid energy density $\ugas$, and the field-parallel velocity $\vpar$, which along with $u^\mu_\perp$ specifies the fluid velocity through \autoref{eq:ugeneral}.

Assuming we know $u^\mu$, to solve for the fluid density $\rho$ we use the advection equation, as in GRMHD:
\begin{equation}
    \nabla_\mu \left(\rho u^\mu\right) = 0.
    \label{eq:densitycons}
\end{equation}
To solve for the fluid energy density, we next assume that the fluid evolves adiabatically in the force-free region:
\begin{align}
    \nabla_\mu \left(\rho s u^\mu\right) &= 0.
    \label{eq:entropycons} 
\end{align}
In \autoref{eq:entropycons}, $s$ is the entropy per unit mass:
\begin{equation}
    s=\frac{1}{\adi-1}\ln\left[\frac{p}{\rho^\adi}\right].
    \label{eq:entropydef}
\end{equation}
The corresponding conserved quantity for the entropy is $\mathcal{S}=s\mathcal{D}$.

The adiabatic approximation \emph{minimizes} the temperature of the fluid in the force-free regions when compared to a solution to the full GRMHD equations. Evolving the fluid adiabatically  will not produce a physical solution to the GRMHD equations in the high $\sigma$ regions, and this choice ignores physical sources of dissipation that may exist in the force-free jet. The standard GRMHD approach of imposing a density floor in the jet tends to produce a jet interior that is relatively high-density and high-temperature; by contrast, the adiabatic GRFFE approach in this region tends to produce an evacuated, lower-temperature funnel (see \autoref{sec:simulations}).


Advecting the density and entropy density with \autoref{eq:densitycons} and \autoref{eq:entropycons} requires us to know both the drift velocity $u^\mu_\perp$ and the parallel velocity $\vpar$ in \autoref{eq:ugeneral}. One option is to keep $\vpar=0$, minimizing the Lorentz factor of the fluid rest frame with respect to the normal observer. However, even when solving for $\vperp$ from pure force-free electrodynamics, we may wish to solve for a nonzero $\vpar$. 
For instance, when coupling GRMHD and GRFFE equations in different regions of the simulation domain, we may have a nonzero $\vpar$ on the boundary of the GRFFE region, so setting $\vpar=0$ in the GRFFE region could introduce discontinuities in the velocity. Furthermore, the decomposition of the velocity into field-parallel and field-aligned components in GRFFE is coordinate-dependent, so zeroing $\vpar$ in the GRFFE region will give different physical velocities when a simulation is run in different coordinate systems.  

An \emph{exact} equation for the evolution of the parallel momentum in GRMHD can be obtained by taking the dot product of the full GRMHD energy-momentum equation $\nabla_\mu T^{\mu\nu}=0$ with the fluid-frame magnetic field 4-vector $b^\mu$ (see \citealt{Camenzind1986} and the Appendix, \autoref{app:pmom} for a derivation):

\begin{align}
    \nabla_\alpha b^\alpha = -\frac{1}{h}b^\alpha\nabla_\alpha p.
    \label{eq:bpargeneral}
\end{align}
\autoref{eq:bpargeneral} indicates that the parallel momentum along a magnetic field line is only changed by gas pressure gradients along the field line. The parallel momentum equation \autoref{eq:bpargeneral} can in principle be used to evolve the parallel velocity $\vpar$, since $b^0=\gamma\vpar\sqrt{\B^2}/\alpha$ by \autoref{eq:b0}.
Unfortunately, the gradient of pressure on the right of \autoref{eq:bpargeneral} makes using \autoref{eq:bpargeneral} to solve for $\vpar$ in an explicit GRMHD code difficult. We can make progress by taking one of two limits. 

In the adiabatic limit, we assume the entropy current is conserved (by \autoref{eq:entropycons}), and additionally that the entropy per unit mass is constant along field lines $\B^i \partial_i s = 0$. Under these assumptions, \autoref{eq:bpargeneral} simplifies to
\begin{equation}
    \nabla_\alpha \left(\mu b^\alpha\right)=0, \;\;\;\; \text{(Adiabatic limit)},
    \label{eq:vparadi}
\end{equation} 
where $\mu$ is the enthalpy-per-unit-mass, $\mu\equiv h/\rho$. 
If we use \autoref{eq:vparadi} together with the GRFFE equations to evolve the parallel velocity, we can obtain $\mu b^0$ as a conserved quantity along with the normal observer frame mass density $\mathcal{D}$ and entropy density $\mathcal{S}$. To invert the system
we then need to numerically solve for $\vpar$ from the conserved quantities $\mu b^0$, and $s=\mathcal{S}/\mathcal{D}$. We can do this numerical inversion using a similar Newton-Raphson method as in standard GRMHD codes \citep{Noble2006}, either iterating on $\vpar$ directly or on $W=\gamma^2h$. 

\autoref{eq:bpargeneral} simplifies further in the cold limit, when we assume $\left|\partial_\alpha p\right| \ll \left|h \nabla_\alpha u^\alpha\right|$. Then 
\begin{equation}
    \nabla_\alpha b^\alpha=0, \;\;\;\; \text{(Cold limit)}
    \label{eq:vparcold}
\end{equation}
In the cold limit, we neglect acceleration of the parallel velocity from pressure gradients along the field line. In this approximation, we can evolve $b^0$ directly as a conserved quantity. Fixing $u^\mu_\perp$ from the force-free equations, we can then exactly solve for $\vpar$ from the known parameter $X^2 \equiv \gamma^2 \vpar^2=\alpha^2 \left[b^0\right]^2 / \B^2$:
\begin{equation}
  \vpar^2 = \frac{\lperp^{-2}X^2}{1+X^2}.
\end{equation}

The cold limit thus provides an explicit prescription for obtaining $\vpar$ by evolving one additional equation for $b^0$ in addition to the force-free solution for $\B^i$ and $u_\perp^i$. Because of its computational simplicity and simple analytic inversion, we adopt the cold limit (\autoref{eq:vparcold}) in this paper. In the case of GRMHD simulations of black hole accretion, using the cold limit of the coupled GRFFE + parallel momentum + adiabatic equations in the high $\sigma$ jet region will not exactly solve the gas dynamics, but it should give a better approximation to the evolution of the gas density, temperature, and momentum in this region than in the standard treatment when density floors are imposed. 

\subsection{Hybrid Evolution}
\begin{figure*}
\centering
\includegraphics[width=.9\textwidth]{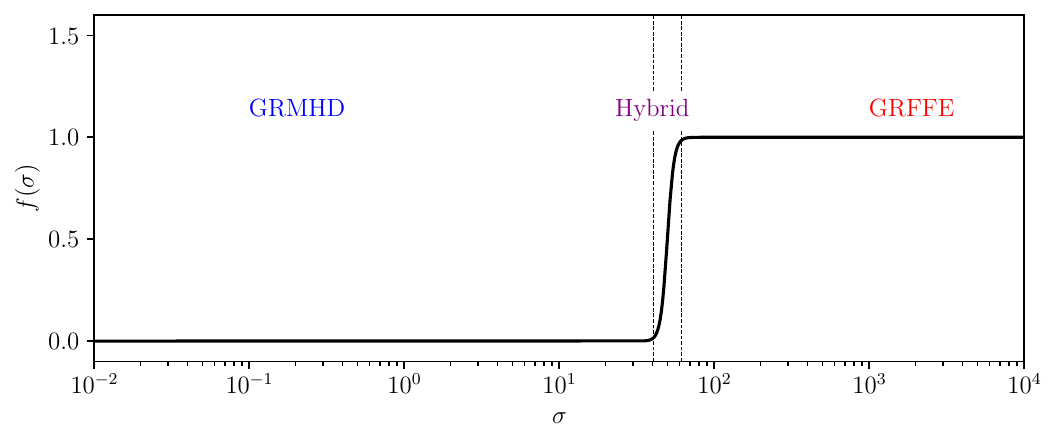}
\caption{
Plot of the mixing function $f(\sigma)$ for the GRMHD simulations in this paper. We center the transition at $\sigma_{\rm trans}=50$, and use width $w=0.1$ in \autoref{eq:fsigma}. Below $f(\sigma)=1/64$ and above $f(\sigma)=63/64$, we switch to entirely GRMHD or GRFFE evolution, respectively. }
\label{fig:fsigma}
\end{figure*}

Finally, we propose a method to couple GRMHD to GRFFE in different regions of a finite volume simulation. In particular, we are interested in switching from solving the full GRMHD equations to solving the approximate system of GRFFE plus the parallel velocity and entropy advection equations described in \autoref{sec:decouple} in
regions of high magnetization $\sigma\gg1$.  This method allows us to evolve the gas density to very small values with $\sigma\gg100$, a typical ``ceiling'' value in GRMHD. 

Our approach is conceptually simple. We keep the same 'primitive' expressions in each cell as in GRMHD (with the addition of the entropy per unit mass $s$):  
\begin{equation}
    \mathbf{P} = \left[\rho,\; \ugas, \; \tilde{u}^i, \;B^i, \;s \right].
    \label{eq:prims}
\end{equation}
In addition to the standard GRMHD conserved quantities (\autoref{eq:conserved}), we add additional ``conserved'' quantities for force-free evolution, corresponding to the three field-perpendicular momenta $\mathcal{Q}_{\mathrm{EM},i} = \alpha T^0_{{\rm EM}, i}$, the conserved quantity for the parallel velocity $\mu b^0$, and the conserved entropy density $\mathcal{S}$:
\begin{align}
    \mathbf{U} &= \frac{\sqrt{-g}}{\alpha} \times \left[\mathcal{D}, \mathcal{Q}_0,  \mathcal{Q}_i,  \B^i,\mathcal{Q}_{\mathrm{EM},i}, \alpha \mu b^0,  \mathcal{S} \right].
     \label{eq:conserved2}
\end{align}
If we use the cold limit for the parallel velocity (\autoref{eq:vparcold}), we set $\mu=1$ in \autoref{eq:conserved2}. We explicitly evolve $\mathbf{U}$ forward in time in every cell using either the GRMHD or force-free equations, as appropriate for the given conserved quantity; $\B^i$ is evolved identically for both sets of equations from the homogeneous Maxwell equation (\autoref{eq:MaxHom}). Then, we determine which quantities to use in the inversion $\mathcal{U}\rightarrow\mathcal{P}$ based on the magnetization $\sigma$ of each cell during the last timestep. 

If the magnetization was less than some critical value $\sigma<\sigma_{\rm trans}$ at the last timestep, we use the standard GRMHD inversion procedure and obtain $\mathbf{P}_{\rm MHD}$. If the magnetization was higher than some critical value  $\sigma>\sigma_{\rm trans}$, we use the force-free conserved quantities to obtain the updated primitives $\mathbf{P}_{\rm FFE}$ following the method in \autoref{sec:grffe} and \autoref{sec:decouple}. 

In practice in a standard GRMHD torus simulation, the vast majority of the simulation cells are in the ``GRMHD'' regime and GRMHD $\mathbf{U}\rightarrow\mathbf{P}$ inversion takes place normally. In some cells concentrated in the simulation jet region, $\sigma\gg1$, and we use the force-free inversion equations.  We turn off density floors and magnetization ceilings entirely in the GRMHD region; whenever these would be needed, we change the equations and conserved quantities in the inversion $\mathbf{U}\rightarrow{\mathbf{P}}$ instead of injecting extra density. 

Instead of switching discretely from one inversion scheme to another depending on $\sigma$, we can allow for a transition between the GRMHD and GRFFE solutions for the inversion by adding a mixing fraction $f(\sigma)$: 
\begin{equation}
    \mathbf{P} = \left(1-f(\sigma)\right)\mathbf{P}_{\rm MHD}(\mathbf{U}) + f\left(\sigma\right)\mathbf{P}_{\rm FF}(\mathbf{U})
    \label{eq:mix}
\end{equation}
We determine $f$ in the range $[0,1]$ by using the local value of $\sigma$ computed at the last time step in a given cell of the simulation. When $f\approx0$ the cell is inverted with GRMHD and when $f\approx1$ it is inverted with the GRFFE conserved quantities. We use a hyperbolic tangent transition function in $\mathrm{ln}\,\sigma$, with a functional form:  
\begin{equation}
f(\sigma) = \frac{\left(\sigma/\sigma_{\rm trans}\right)^{2/w}}{1+\left(\sigma/\sigma_{\rm trans}\right)^{2/w}}.
\label{eq:fsigma}
\end{equation}
In \autoref{eq:fsigma}, the parameter $\sigma_{\rm trans}$ is the value of $\sigma$ where the transition from GRMHD to GRFFE is centered, and $w$ controls the width of the transition. The limit $w\rightarrow0$ corresponds to an sharp transition exactly at $\sigma_{\rm trans}$.

Using finite transition width between the GRMHD and GRFFE solutions for $\mathbf{P}(\mathbf{U})$ in \autoref{eq:mix} has the disadvantage of not exactly conserving energy and momentum under either the GRMHD or GRFFE equations in the transition region. However, it may be useful to smooth out any sharp discontinuities in the simulation from the boundary between the two regions. In practice, in GRMHD torus simulations, we find that $\sigma$ climbs so rapidly in the transition into the jet region that only a few cells along the jet sheath are not in either the limit $f\approx 0$ or $f\approx 1$. 

For computational efficiency, and to ensure the bulk of the simulation either purely GRMHD or purely GRFFE, we use only GRMHD inversion when $f<1/64$ and only force-free inversion when $f>63/64$. This amounts to mixing the results on the two inversion procedures only between $\sigma_{\rm low}=\sigma_{\rm trans}/ f_c$ and $\sigma_{\rm high} = \sigma_{\rm trans} f_c$, where $f_c=3^w 7^{w/2}$. We plot a typical curve $f(\sigma)$ with $\sigma_{\rm trans}=50$ and $w=0.1$ in \autoref{fig:fsigma}.

\section{Implementation and Code Tests}
\label{sec:tests} 

Here we present details of our implementation of both GRFFE and hybrid GRMHD+GRFFE evolution in the code \texttt{KORAL} \citep{KORAL13}. We then present a number of 1D and 2D code tests of both pure GRFFE and GRMHD+GRFFE hybrid problems in both flat space and the Kerr geometry.

\subsection{Implementation Details}

We have implemented both the GRFFE solution method of \citet{McKinney06} and its extension to hybrid GRFFE+GRMHD evolution in the GRMHD code \texttt{KORAL} \citep{KORAL13}. \texttt{KORAL} was based on the \texttt{HARM} GRMHD code \citep{Gammie03,Noble2006}, but is extensively modified for radiative GRMHD simulations of black hole accretion \citep{KORAL14,KORAL15} as well as two-temperature radiative simulations \citep{KORAL16,Chael17}. The \texttt{KORAL} code has been extensively tested on standard GRMHD test problems \citep{KORAL14} and compared to other GRMHD codes in the EHT code comparison project  \citep{Porth2019}.

\texttt{KORAL} works on a regular grid in arbitrary spacetime coordinates; all quantities (including the magnetic field) are cell-centered. In most applications, \texttt{KORAL} interpolates primitives $\mathbf{P}$ to cell walls with the second order piecewise parabolic method (PPM). Fluxes at cell walls are evaluated with the local Lax-Friedrichs (LLF) method, and evolution in time is performed with a second-order Runge-Kutta method. \texttt{KORAL} ensures the magnetic field remains divergence-free by using the FluxCT constrained transport algorithm \citep{Toth2000}. 

In standard GRMHD applications, \texttt{KORAL} applies ceilings on the magnetization $\sigma$ in the normal observer frame, but an option to apply ceilings in the drift velocity frame \citep{Ressler17} is also implemented. A typical value of the ceiling magnetization value used in GRMHD evolution in KORAL is $\sigma_{\rm max}\approx 100$. In the GRFFE and hybrid GRMHD+GRFFE code tests presented here, we completely turn off the standard  $\sigma_{\rm max}$ ceiling in \texttt{KORAL} and instead transition to force-free evolution at a given $\sigma_{\rm trans}$, using either a sharp transition or applying a smoothing function $f(\sigma)$, as in \autoref{eq:mix}. In hybrid GRMHD+GRFFE evolution, we update the local $f(\sigma)$ of a cell once per Runge-Kutta sub-timestep. As discussed above, when using a continuous mixing function $f(\sigma)$, we apply upper and lower limits on $\sigma$, $\sigma_{\rm high}$ and  $\sigma_{\rm low}$,  above and below which quantities are evolved only according to the GRFFE or GRMHD equations, respectively.
In all applications we must retain a ceiling on the maximum Lorentz factor; we typically set $\gamma_{\rm max}=100$ unless otherwise specified.

\subsection{GRFFE Tests}
\begin{figure*}
\centering
\includegraphics[width=.9\textwidth]{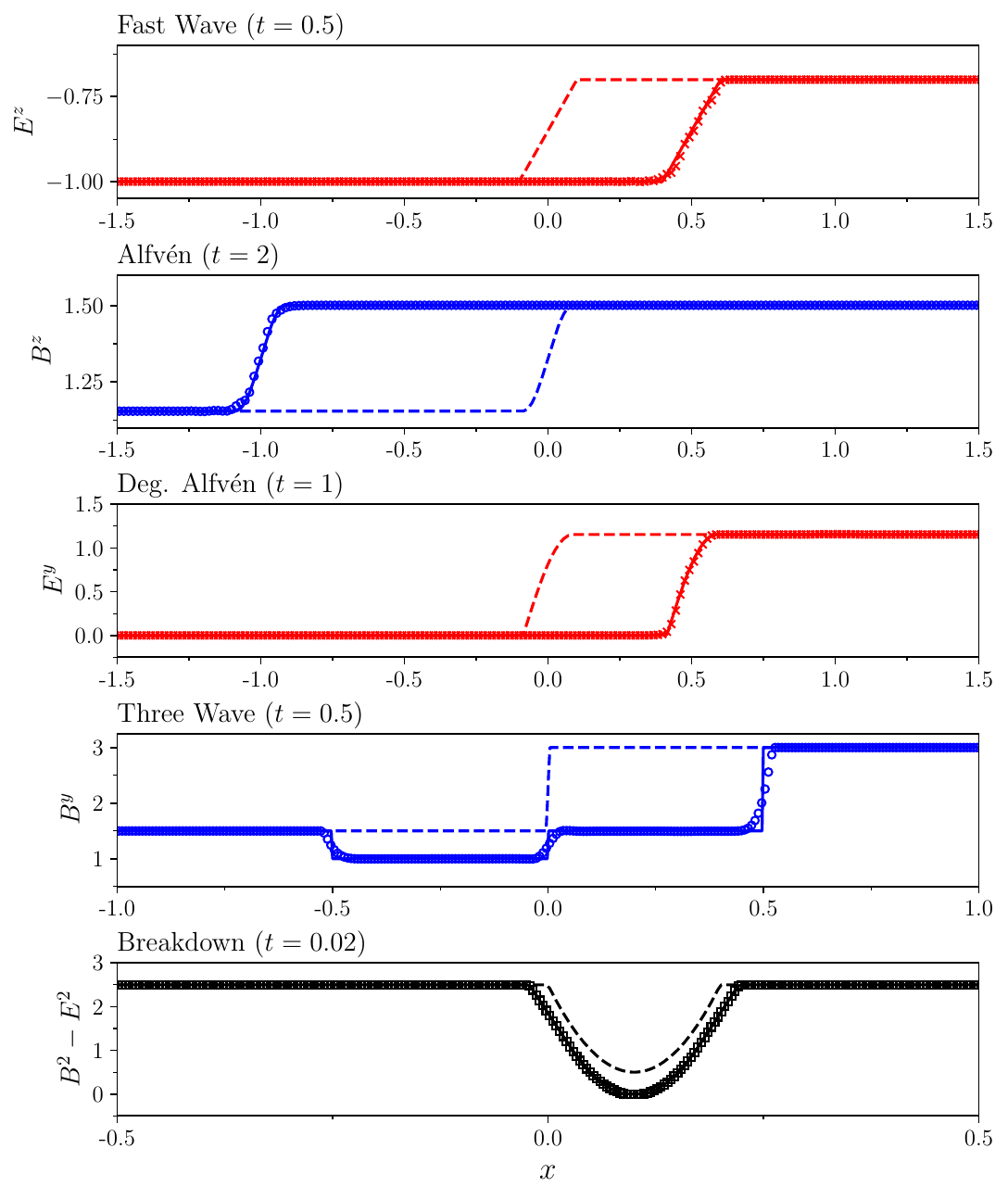}
\caption{
Special relativistic FFE tests from \citet{Komissarov02b}. For each problem, dashed lines represent the initial solution, solid lines indicate the analytic solution at the given time, and symbols indicate the numerical solution for the reported magnetic or electric field component. In the breakdown test (bottom row), there is no analytic result, but our solution closely matches the result from \citet{Komissarov02b} indicating that that at time $t=0.02$ the electric field strength is nearly equal to the magnetic field strength. 
}
\label{fig:komissarovtest}
\end{figure*}

We first test our implementation of pure GRFFE evolution in \texttt{KORAL} with the standard set of 1D test problems in flat spacetime introduced in \citet{Komissarov02b}. These tests consist of (1) a nonlinear fast wave propagating to the right at the speed of light, $v=1$; (2) an nonlinear Alfv\'en wave propagating left with speed $v=-0.5$; (3) a nonlinear degenerate Alfv\'en wave propagating to the right with speed $v=0.5$; (4) a superposition of a stationary Alfv\'en wave and two nonlinear fast waves propagating left and right at $v=1$; (5) a ``breakdown'' test that evolves to a state where the field becomes electrically dominated. See \citet{Paschalidis13} for clear and detailed implementations of the initial conditions in all five problems.

The spatial domain is $x\in[-2,2]$ for the fast wave and Alfv\'en tests, $x\in[-1,1]$ for the three wave test, and $x\in[-0.5,0.5]$ for the breakdown test.  We used outflow boundary conditions in all cases, and a fiducial resolution of $N_x=256$ in all tests. We ran all tests to a maximum time of $t_{\rm max}=1.5$.

Because these tests are pure FFE problems, we turned the parallel momentum solver off and only evolved the drift velocity $u^\mu_{\perp}$ and magnetic field $\B^\mu$. The initial density and internal energy are advected adiabatically, but do not back-react on the magnetic field and velocity. We used first-order spatial reconstruction with a monotonized central (MC) limiter instead of second-order PPM reconstruction, as first-order reconstruction is often more robust for propagating sharp discontinuities \citep{Gammie03}. 

We present our results for all five test problems in \autoref{fig:komissarovtest}; these results may be compared to results from other GRFFE codes, including Figure 1 of \citet{McKinney06} and Figure 1 of \citet{Paschalidis13}. For the fast wave, Alfv\'en and three-wave tests, we plot a given component of the electric or magnetic field evolved by \texttt{KORAL} over the analytic solution at a given time; in all cases, the numerical solution lines up well with the analytic expectation. In the breakdown test (which does not have an analytic solution) we plot $B^2-E^2$, indicating the degree of magnetic domination of the electromagnetic field. As reported by \citet{Komissarov02b}, we see that for the given initial condition the electric field strength approaches the magnetic field strength at $t\approx0.02$, after which the FFE solver must begin applying a ceiling on the Lorentz factor $\gamma$.

\subsection{Hybrid GRMHD+GRFFE tests}
Next we turn to testing our implementation of matching the GRFFE solver in part of the spatial domain with standard GRMHD in the rest of the domain. For a GRMHD problem (like the Bondi problem in \autoref{sec:bondi}),  the GRFFE solution will only be approximate and should match the expected GRMHD solution only in regions of high $\sigma$; similarly, for a GRFFE problem setup, we expect any GRMHD solution to diverge from the GRFFE solution in regions of low $\sigma$. Thus, how well any hybrid solution matches an analytic result will depend on the choice of $\sigma_{\rm trans}$ and desired error tolerance. In these tests, we primarily aim to show that our implementation of hybrid GRMHD+GRFFE does not introduce unexpected artifacts at the boundary between the GRFFE and GRMHD domains. 

\subsubsection{Linear Waves}
\begin{figure*}
\centering
\includegraphics[width=.9\textwidth]{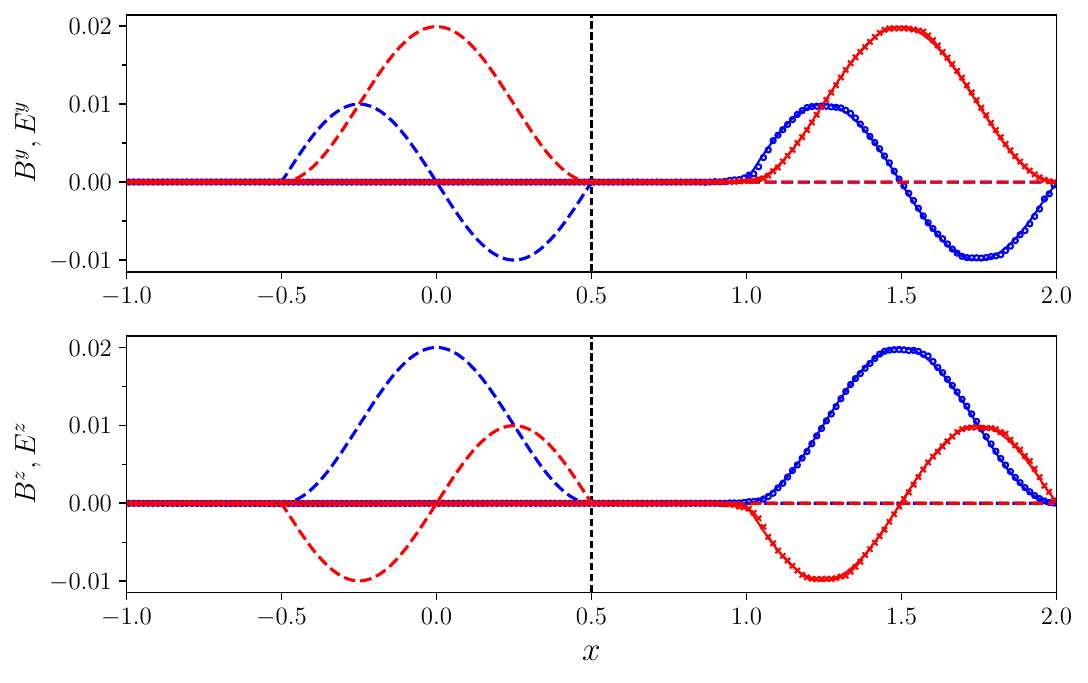}
\caption{
Linear Alfv\'en pulse test. In the top row, we show the $y-$ components of the magnetic and electric field in blue and red, respectively; in the bottom row, we show the $z-$ components in the same colors. Dashed lines indicate the initial condition; the solid lines indicate the analytic expectation for the solution at $t=1.5$ and the open markers show the numerical solution. The numerical solution transitions from standard MHD inversion to the left of the dashed vertical line at $x=0.5$ to FFE inversion to the right of $x=0.5$, with the transition happening over a finite width using the mixing prescription in \autoref{eq:mix}, replacing $f(\sigma)$ with $f(x)$ from \autoref{eq:xmix}.  
}
\label{fig:alf}
\end{figure*}

We first consider an Alfv\'en wave in flat space in the linear regime, moving rightward from a region where we solve the MHD equations in a region of strong magnetic field into a region where we solve the approximate set of FFE and decoupled parallel momentum, density and entropy equations. For our initial conditions, we fix a background B-field $B_0=1$ along the $x-$direction. We fix the magnetization $\sigma=250$, so $\rho=B_0^2/\sigma=0.4$, and we fix the initial temperature $\Theta_{\rm gas}=0.2$ and adiabatic index $\adi=4/3$.

Alfv\'en waves propagating on this MHD background have a velocity $v_{A}=\sqrt{\sigma_w/(1+\sigma_w)}=0.996$, where $\sigma_w = b^2/h=138.\bar{8}$ is the magnetization defined relative to the enthalpy. Our initial pulse propagates to the right with this velocity in the MHD region and at the speed of light in the FFE region. We define our initial pulse in the domain $-0.5<x<0.5$: 
\begin{align}
B^x &= B_0 \nonumber \\
B^y &= -\epsilon B_0 \sin kx \nonumber \\
B^z &= \epsilon B_0 \left(1+\cos kx\right),
\end{align}
while for $x>0.5$ and $x<-0.5$, $B^x=B_0$ and $B^y=B^z=0$. We fix $\epsilon=10^{-3}$. The drift velocity of the initial pulse is 
\begin{align}
v^x &= 0 \nonumber \\
v^y &= -v_A B^y/B_0 \nonumber \\
v^z &= -v_A B^z/B_0.
\end{align}

Rather than determine which solver we use based on the local $\sigma$, in this problem we transition from GRMHD to GRFFE based on the spatial coordinate $x$. We thus replace $f(\sigma)$ in \autoref{eq:mix} with $f(x)$ defined by
\begin{equation}
\label{eq:xmix}
f(x) = \frac{1}{2} + \frac{1}{2}\mathrm{tanh}\left(\frac{x-x_{\rm trans}}{w}\right),
\end{equation}
where $x_{\rm trans}=0.5$ and $w=0.1$. Again, we use MHD inversion exclusively when $f<1/64$ and FFE inversion exclusively when $f>63/64$. Our spatial domain is $x\in[-2,2]$, and we use a fiducial resolution of $N_x=512$. We run the simulation to a maximum time of $t_{\rm max}=1.5$. In the FFE solution, we use the cold approximation to determine the parallel momentum. We use second-order PPM spatial reconstruction, and limit our maximum Lorentz factor to $\gamma<\gamma_{\rm max}=1000$.

We present our results in \autoref{fig:alf}, where we plot the $y-$ and $z-$ components of the electric and magnetic field (orthogonal to the background magnetic field in the $x-$direction) after evolving the system to $t_{\rm max}=1.5$. The numerical solution lines up well with the analytic expectation of a propagating linear pulse and does not show any artifacts from passing through the transition between the MHD inversion and FFE inversion at $x=0.5$. 

\subsubsection{Bondi Accretion}
\label{sec:bondi}

\begin{figure*}
\centering
\includegraphics[width=.8\textwidth]{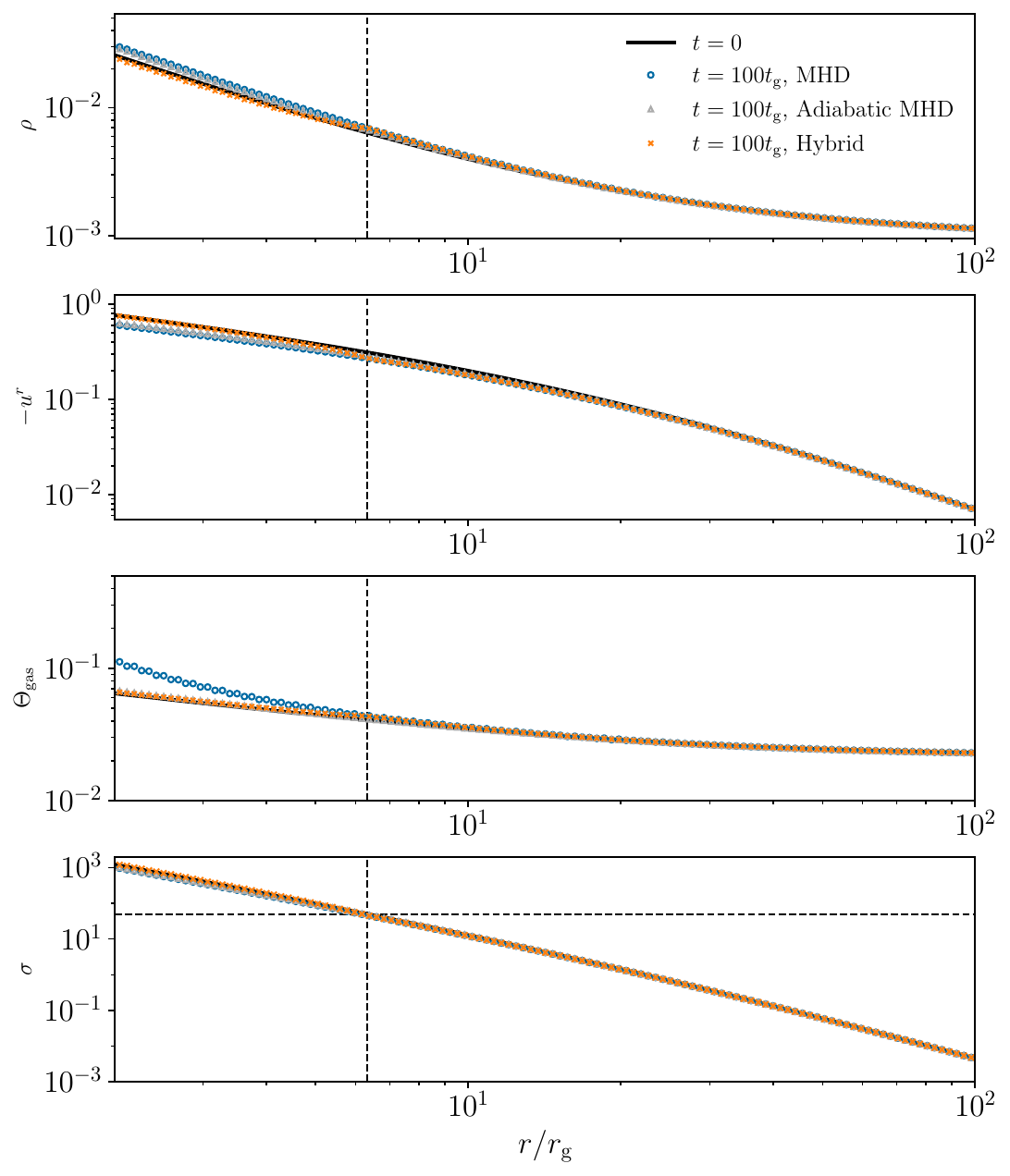}
\caption{
Schwarzschild Bondi Accretion Test Problem. From top to bottom, we show profiles of the plasma density $\rho$, Boyer-Lindquist radial velocity $u^r$, dimensionless temperature $\Theta_{\rm gas}$, and magnetization $\sigma$ for the spherically symmetric Bondi accretion setup described in this section. The solid black line indicates the initial condition; the Bondi problem should remain stationary in time. Blue circles show the numerical solution after $100\,t_{\rm g}$ for a standard GRMHD method, grey triangles show the results from a GRMHD simulation which evolves the fluid energy adiabatically, and orange xs show the numerical solution at the same time from our hybrid approach. In the hybrid approach, we transition between GRMHD and GRFFE solutions at $\sigma_{\rm trans}=50$, indicated by the horizontal dashed line in the bottom panel; this transition magnetization corresponds to a stable transition radius of $r\approx 6.3 \,r_{\rm g}$, indicated by the vertical dashed line in all panels. 
}
\label{fig:bondi}
\end{figure*}

\begin{figure*}
\centering
\includegraphics[width=.9\textwidth]{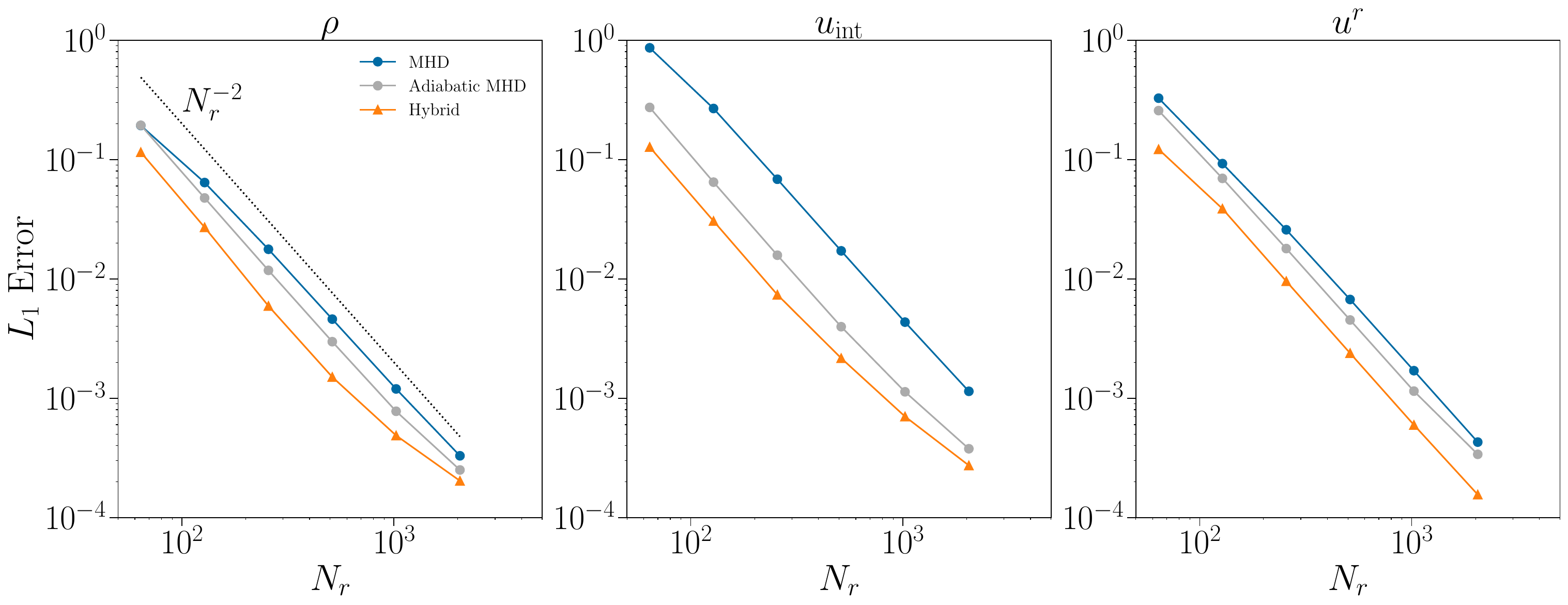}
\caption{
Convergence test for the Bondi problem. From left to right, we plot the $L_1$ error (\autoref{eq:l1}) between the solution at $t=100\,t_{\rm g}$ and the analytic solution for the density $\rho$, internal energy $\ugas$, and Boyer-Lindquist radial velocity $u^r$ as a function of the number of radial cells $N_r$. We plot the $L_1$ error for the standard MHD run in blue circles and the error for the hybrid GRMHD+GRFFE run in orange triangles. Results from the adiabatic GRMHD simulation are shown in grey. 
}
\label{fig:bondiconv}
\end{figure*}

In this test we simulate spherically symmetric Bondi accretion onto a Schwarzschild ($a_*=0$) black hole, which is a standard test problem in GRMHD codes \citep[e.g.][]{DeVilliers03b,Bhac}. We set up an analytic Bondi flow in Boyer-Lindquist coordinates with a sonic point at $r_{\rm c}=8 \,r_{\rm g}$, an accretion rate $\dot{M}=-1$ in code units, and adiabatic index $\adi=4/3$. We then solve for the initial density, internal energy, and velocity for the Bondi solution as a function of radius in the equatorial plane numerically following the description in \citet{RezzollaBook}, section 11.4. 

A magnetic field parallel to the radial infall does not affect the Bondi solution \citep{Bhac}. We set up an initial radial magnetic field,
\begin{equation}
B^r = \frac{\sigma_{\rm c}}{\rho_{\rm c}}\left(\frac{r}{r_{\rm c}}\right)^{-2}
\end{equation}
Where the velocity and density at the sonic point are
\begin{align}
u^r_{\rm c} &= -\frac{1}{2 r_{\rm c}} \\
\rho_{\rm c} &=\frac{\dot{M}}{4\pi r^2_{\rm c} u^r_{\rm c}}
\end{align}
We fix the magnetization at the sonic point to be $\sigma_{\rm c}=25$. The magnetization at the horizon climbs to $\sigma\approx1000$. 

The Bondi solution should remain stationary in time.
We evolve the initial conditions numerically in 1D in modified Kerr-Schild (KS) coordinates in the domain $r/r_{\rm g}\in\left[1.8,100\right]$. We use a logarithmic grid in $r$; the resolution is $N_r=128$ in our fiducial test. 
While we run the simulation in KS coordinates, we present our results in Boyer-Lindquist coordinates in \autoref{fig:bondi}. We use outflow boundary conditions and second order PPM reconstruction, and we run the problem to $t_{\rm max}=100 \,t_{\rm g}$. 

In \autoref{fig:bondi} we compare results from our new hybrid GRMHD+GRFFE method to a standard GRMHD run of the same problem using \texttt{KORAL}. When testing the hybrid method, we fixed the transition point at $\sigma_{\rm trans}=50$ in \autoref{eq:fsigma}, with width $w=0.05$. This choice puts the transition around $r\approx6.3\,r_{\rm g}$ in our problem set-up, indicated by the vertical dashed line in \autoref{fig:bondi}.  
For the given set-up, standard GRMHD is well behaved even up to $\sigma=1000$ at the horizon, so we we do not implement any ceiling on $\sigma$ in the GRMHD comparison run. 

This problem tests the parallel momentum solver in the force-free region, as all of velocity is parallel to the radial magnetic field line and hence $u^\mu_\perp=0$. Because the dimensionless temperature is relatively high close to the horizon, $\Theta_{\rm gas}\approx0.1$, we use the adiabatic approximation (\autoref{eq:vparadi}) instead of the cold approximation (\autoref{eq:vparcold}) for solving for the parallel velocity in the force-free region. 

In \autoref{fig:bondi} we show the initial conditions and the GRMHD and hybrid GRMHD+GRFFE solutions after evolving the system for $t=100\,t_{\rm g}$ for the plasma density $\rho$, radial four-velocity $u^r$, temperature $\Theta_{\rm gas}$ and magnetization $\sigma$ from our fiducial run with resolution $N_r=128$. The standard MHD solution begins to deviate from the analytic stationary solution close to the horizon, as $\sigma\gsim100$; by contrast, the hybrid solution more closely tracks the analytic expectation in this region as the code switches from solving the GRMHD to GRFFE equations, including the approximate equations for the parallel velocity and adiabatic internal energy evolution. In particular, the MHD solution under-predicts the infall velocity and over-predicts the plasma temperature $\Theta_{\rm g}$ in the high magnetization region close to the horizon, while the hybrid solution more closely tracks the ground truth. We also show results from an GRMHD run where the conserved entropy $\mathcal{S}$ is used instead of the energy $\mathcal{Q}_0$; when adiabatic evolution is enforced, the standard GRMHD run recovers the gas temperature significantly more accurately \citep{Bhac}, but it still performs worse than the hybrid run in recovering $\rho$ and $u^r$. 

In \autoref{fig:bondiconv} we perform a convergence test for both methods. We plot the normalized $L_1$ error for the density $\rho$, internal energy $u_{\rm int}$ and velocity $u^r$ of the numerical solution at $t=100\,t_{\rm g}$ compared to the stationary ground truth initial condition. The $L_1$ error for a given quantity $q(r,t)$ is defined,
\begin{equation}
\label{eq:l1}
L_1(q,t)= \frac{1}{N_r} \sum_i \frac{\left|q(r_i,t)-q(r_i,0)\right|}{\left|q(r_i,0)\right|}.
\end{equation}
where we sum over all radial points $r_i$ in the domain. \autoref{fig:bondiconv} indicates that the accuracy of both the standard GRMHD and hybrid  GRMHD+GRFFE approaches converges at second order in the spatial resolution, but the absolute error of the hybrid set-up is lower for all three quantities than the MHD solution. As seen in \autoref{fig:bondi}, the standard MHD approach introduces particularly large error in the internal energy $u_{\rm int}$ when compared to the error in the density or velocity. By contrast, the hybrid approach features a similar magnitude of the relative error in all three quantities for a given resolution $N_r$. The GRMHD simulation run with adiabatic evolution enforced has an absolute error in all three quantities that is between the standard GRMHD and hybrid simulations.  

The GRMHD solution is stable at the given magnetization $\sigma_{\rm c}=25$ and does not need to impose ceilings on the magnetization close to the horizon. However, if we increase the magnetization further beyond $\sigma_{\rm c}=100$, standard GRMHD begins to break down close to the horizon, while our new hybrid method remains well-behaved. 

\subsubsection{BZ Monopole}
\label{sec:monopole}
\begin{figure*}
\centering
\includegraphics[width=.8\textwidth]{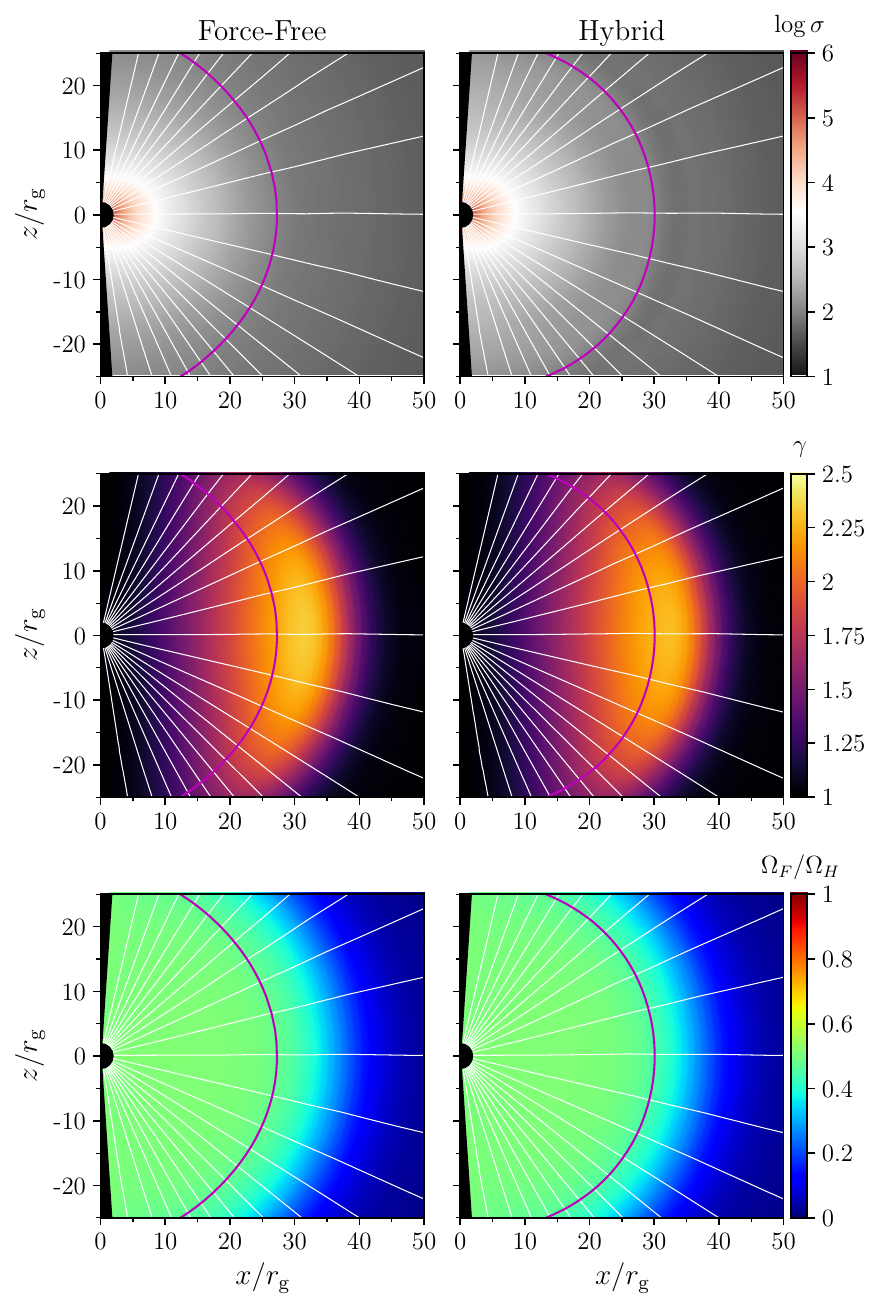}
\caption{
2D BZ Monopole Test. From top to bottom, we show poloidal profiles of the magnetization $\sigma$, Lorentz factor $\gamma$ in KS coordinates, and fieldline angular speed $\Omega_F/\Omega_H$ for the monopole simulations presented in this section with dimensionless black hole spin $a_*=0.5$. We show results from the end of the simulation at $t=50t_{\rm g}$. In the left column, we plot results from a purely force-free simulation, and in the right column we show results from a hybrid GRMHD+GRFFE simulation with the same initial conditions. The magenta contour in all plots shows the $\sigma=100$ surface, which is the center of the transition function $\sigma_{\rm trans}=100$ in the hybrid simulation. }
\label{fig:monopole}
\end{figure*}

\begin{figure*}
\centering
\includegraphics[width=.9\textwidth]{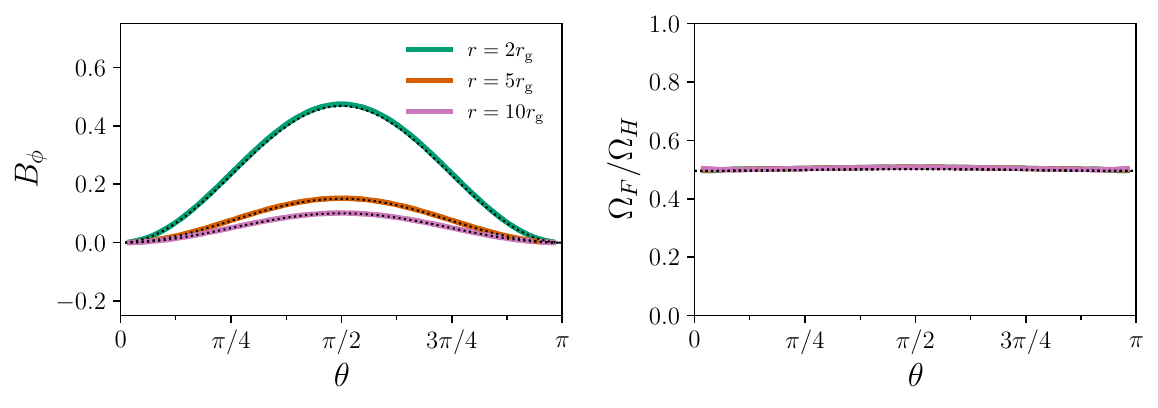}
\caption{
Simulation comparison to the BZ monopole analytic solution. In the left plot, we show angular slices of the covariant toroidal magnetic field $B_\phi=g_{i\phi}\Fd^{i0}$ at radii $r=2\,r_{\rm g}$ (grey), $r=5\,r_{\rm g}$ (blue), and  $r=10\,r_{\rm g}$ (orange) for the hybrid monopole simulation shown in \autoref{fig:monopole} at $t=50\,t_{\rm g}$. In the right panel, we show profiles of the fieldline angular frequency $\Omega_F$ at the same radii, normalized by the horizon angular frequency $\Omega_H$. The analytic solution from \citet{BZ} for each radius is plotted in a dotted line.
}
\label{fig:monopoleprof}
\end{figure*}

For our final test problem we simulate a 2D force-free monopole in the Kerr metric for a black hole with dimensionless spin $a_*=0.5$. We set up an initial radial monopolar field in the Kerr spacetime. As the field evolves under the GRFFE equations, it is quickly spun up by the black hole and develops a significant toroidal component. The wound-up magnetic field achieves an angular velocity $\Omega_F=0.5\Omega_{H}$, where $\Omega_{H}=a/2Mr_+$ is the angular frequency of the horizon, thus maximally extracting energy from the black hole spin by the \citet{BZ} process. The BZ monopole is a standard test problem for GRFFE codes \citep[e.g.][]{Komissarov02b, McKinney06,Paschalidis13, Mahlmann2021}; it can also be evolved to relatively high $\sigma$ in GRMHD codes \citep[e.g.][]{Tchekhovskoy2009}. 

We work in Kerr-Schild coordinates $(t,r,\theta,\phi)$ in two dimensions. Our initial monopolar magnetic field is set by a toroidal vector potential $A_\phi$: 
\begin{equation}
A_\phi=1-\cos\theta.
\end{equation}
The initial radial magnetic field in the normal observer frame is $\B^r = \alpha \partial_\theta A_\phi /\sqrt{-g}$. 
We set up an initial density profile $\rho(r)$:
\begin{equation}
\rho(r) = \frac{B^2}{\sigma_{\rm init}}\left(\frac{r}{r_+}\right)^{-3},
\end{equation}
where $\sigma_{\rm init}=1000$. The atmosphere is initially isothermal with dimensionless temperature $\Theta_{\rm gas}=10^{-4}$, and the adiabatic index is $\adi=4/3$. The initial velocities in the normal observer frame are all zero.  

We evolve the problem forward in time on a regular 2D grid in modified Kerr-Schild coordinates $(t,x_1,x_2,\phi)$, defined in \texttt{KORAL} by
\begin{align}
\label{eq:mks}
    r&=e^{x_1}+r_0 \nonumber \\
    \theta&=\frac{\pi}{2}\left[1+\cot\left(\frac{\pi}{2}\right)\tan\left(\frac{\pi}{2}h_0\left(2x_2-1\right)\right)\right].
\end{align}
For this test we use $r_0=0$ and $h_0=0.8$. We use $N_r=192$ radial cells in the range $r\in \left[0.7r_+, 250\,r_{\rm g}\right]$ and $N_\theta=128$ cells in polar angle between $x_2\in[0.001,0.999]$. We run the simulation to a maximum time $t_{\rm max}=50\,t_{\rm g}$. 

For this problem, we compare a simulation run with pure GRFFE to a hybrid simulation that transitions from GRFFE to GRMHD evolution at $\sigma_c=100$, with a width $w=0.1$ in \autoref{eq:fsigma}. In the GRMHD region of the hybrid simulation, we enforce entropy conservation (replacing $\mathcal{Q}^0$ with $\mathcal{S}$ in the GRMHD conserved-to-primitive inversion, \autoref{app:u2p}).  In both simulations, in the FFE region we solve for the field-parallel momentum under the cold approximation, \autoref{eq:vparcold}.
We found that using the parallel momentum solver is essential even in the inner GRFFE region of the hybrid simulation to correctly match the boundary conditions set by GRMHD evolution in the outer region. In both simulations we use outflow boundary conditions and first-order spatial reconstruction with monotonized central (MC) limiter. We set the maximum Lorentz factor $\gamma_{\rm max}=2000$.

As the monopolar field winds up and begins extracting energy from the black hole, it launches an outflow, driving plasma from the central region and expanding the transition radius where $\sigma=\sigma_{\rm trans}=100$. In  \autoref{fig:monopole} we show 2D poloidal slices of the magnetization $\sigma$, normal observer Lorentz factor $\gamma$ in KS coordinates, and field-line angular frequency $\Omega_F$ for both the GRFFE and hybrid GRFFE+GRMHD simulations. For an axisymmetric degenerate electromagnetic field, the fieldline angular frequency can be computed from the plasma velocity and magnetic field components as
\begin{equation}
\Omega_F = \frac{u^\phi}{u^t} -\frac{u^r}{u^t}\frac{\B^\phi}{\B^r}.
\end{equation}

The results shown in \autoref{fig:monopole} are not identical between the completely force-free simulation and hybrid GRMHD+GRFFE simulation. The differences are most pronounced at the transition region between GRMHD and GRFFE evolution at the $\sigma_{\rm trans}=100$ contour, which slightly lags the outer radius where the field lines are wound up by the black hole (the outer radius with $\Omega_F\approx0.5\Omega_H$).  The pure force-free simulation achieves a slightly higher Lorentz factor than the hybrid simulation, and the $\sigma=100$ contour is at a slightly smaller radius in the force-free simulation at $t=50t_{\rm g}$. Compared to the force-free simulation, the hybrid simulation has a slightly $(\lsim 5\%)$ smaller radial velocity at and just outside the transition surface but nearly the same radial velocity deeper in the force-free region; as a result, some gas piles up outside the expanding wind front in the hybrid simulation, slightly lowering $\sigma$ compared to the pure force-free result.

Despite these differences, the overall structure of all three quantities in both simulations is similar, particularly interior to the transition radius in the force-free region at $r\approx25\,r_{\rm g}$. Thus, the boundary between GRMHD evolution and GRFFE evolution in our hybrid scheme does not adversely affect the force-free evolution inside the expanding region of wound-up field lines in this set-up. 

In \autoref{fig:monopoleprof} we show angular profiles of the poloidal covariant toroidal field $B_\phi$ and field line angular speed $\Omega_{F}$ for the hybrid simulation compared to the analytic \citet{BZ} solution at different radii. The simulation results closely track the analytic solution; in particular, the fieldline angular speed achieves the expected optimal value for energy extraction, $\Omega_F=0.5\Omega_H$. 

\section{First 3D MAD Simulations in GRMHD and Hybrid GRMHD+GRFFE}
\label{sec:simulations}

\begin{table*}
    \centering
    \begin{tabular}{l|ccc|cc|cc|cc|cc}
        \toprule 
        Model  & $a_*$ & $\adi$ & $B$-field & $r_{\rm in}/r_{\rm g}$ & $r_{p,\mathrm{max}}/r_{\rm g}$& $N_r \times N_\theta \times N_\phi$ & $\left[r_{\rm min}/r_{\rm g}, r_{\rm max}/r_{\rm g}\right]$ & $\sigma_{\rm trans}$ & $\gamma_{\rm max}$ & $\langle \dot{M} \rangle$ & $\langle \phi \rangle$ \\
        \midrule
        MHD &  0.5 & $13/9$& MAD & 20  & 42 & $160\times128\times96$  & $\left[1.5,1000\right]$ & 50 & 100 & 45.2 & 52.8 \\
        Hybrid &  0.5 & $13/9$& MAD & 20  & 42 & $160\times128\times96$  & $\left[1.5,1000\right]$ & 50 & 100 & 37.1 & 56.4  \\
        \bottomrule
    \end{tabular}
    \caption{Summary of 3D MAD simulations presented in this work. The mass accretion rate through the horizon $\dot{M}$ and the normalized magnetic flux $\phi$ were averaged over the range $[5000,10000]$ $t/t_{\rm g}$.
    }
    \label{tab:model_summary}
\end{table*}

\begin{figure*}
\centering
\includegraphics[width=.8\textwidth]{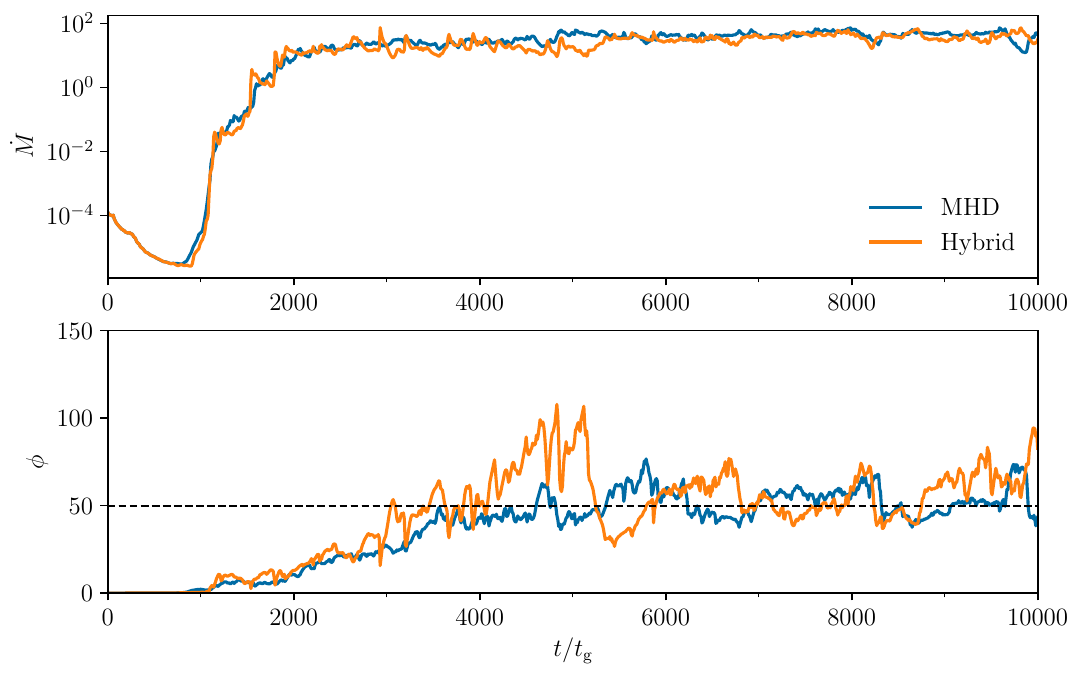}
\caption{
Evolution of the mass accretion rate $\dot{M}$ (top) and dimensionless magnetic flux through the horizon $\phi$ (bottom; \autoref{eq:madpar}) in the GRMHD and Hybrid MAD simulations from $t=0$ to $t=10^4t_{\rm g}$. 
}
\label{fig:scalars}
\end{figure*}

\begin{figure*}
\centering
\includegraphics[width=.8\textwidth]{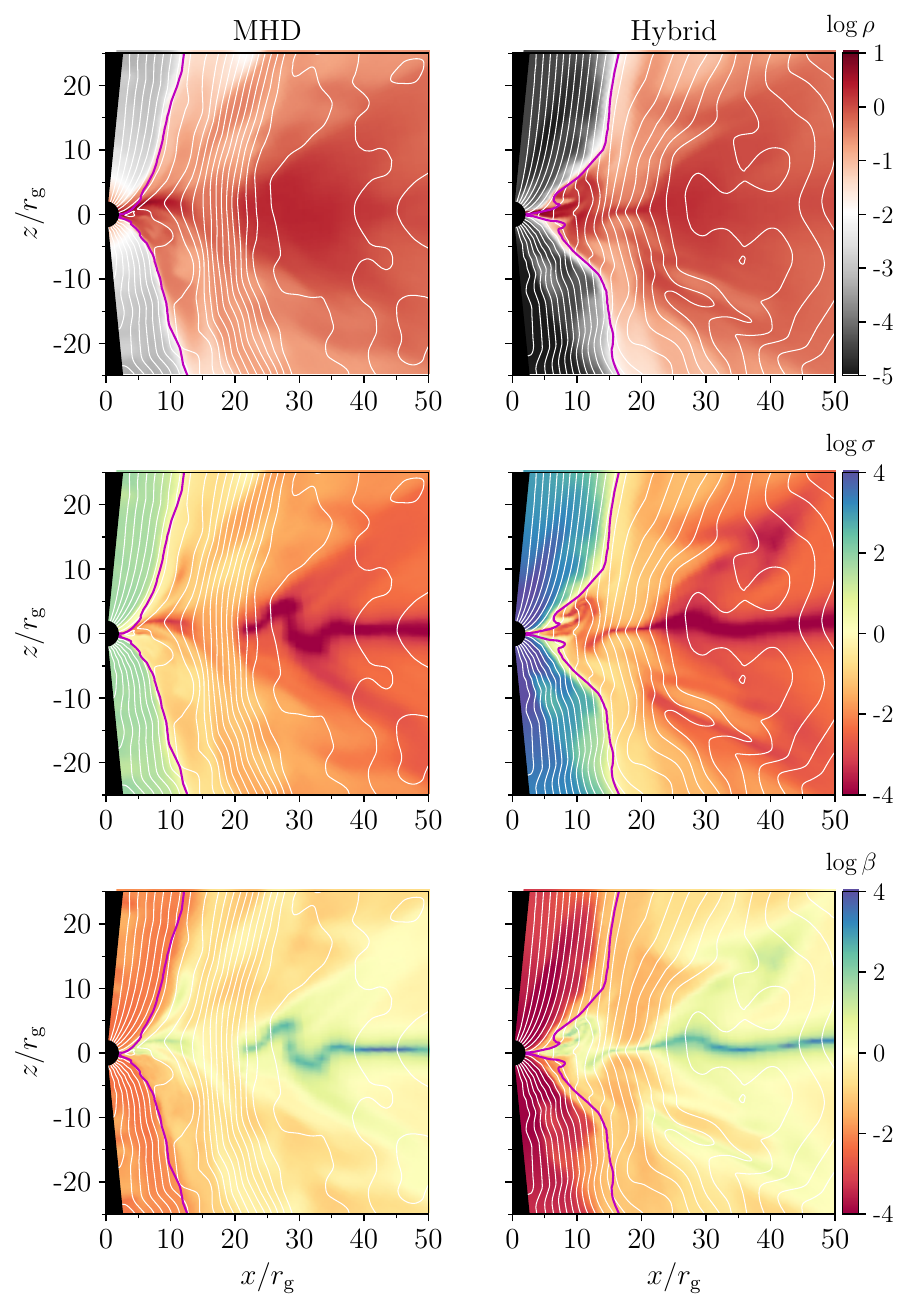}
\caption{
Snapshot poloidal slices from the GRMHD (left column) and Hybrid GRMHD+GRFFE (right column) MAD simulations. From top to bottom, we plot the plasma mass density $\rho$ in code units; the magnetization $\sigma$; and the ratio of thermal to magnetic pressure $\beta$. Both simulation snapshots were taken at $t=8550\,t_{\rm g}$. In all plots we present snapshot data from a slice of constant azimuthal angle $\phi$ in Kerr-Schild coordinates. Contours of constant $A_\phi$, the azimuthal component of the vector potential, are shown in white. The black disc indicates the black hole horizon, and the magenta contour in all plots indicates the $\sigma=1$ surface.
}
\label{fig:phisli}
\end{figure*}

\begin{figure*}
\centering
\includegraphics[width=.8\textwidth]{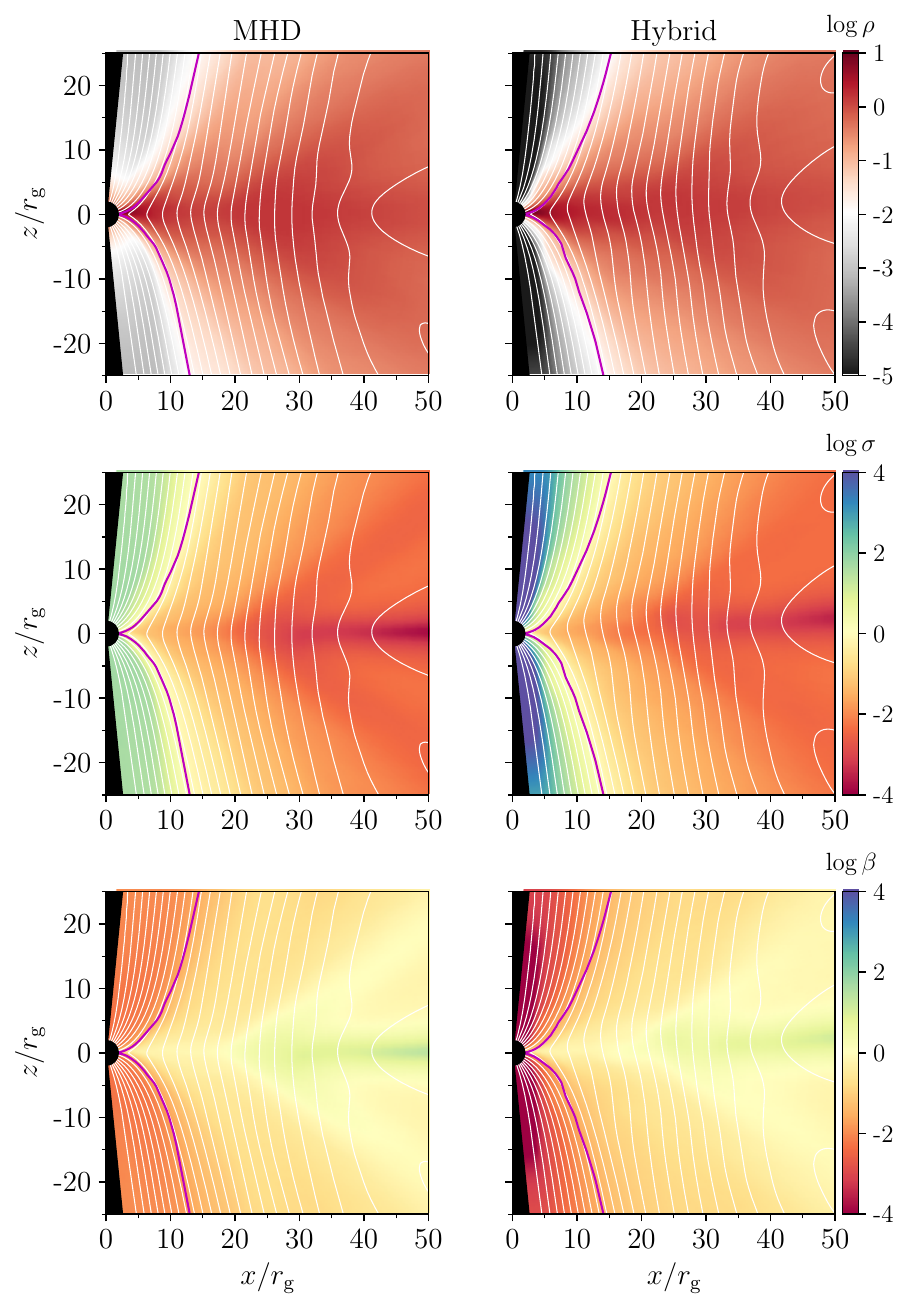}
\caption{
Time- and azimuth-averaged poloidal profiles of the mass density $\rho$ (top; in code units), the magnetization $\sigma$, and the ratio of thermal to magnetic pressure $\beta$. Both the standard MHD simulation data (left column) and hybrid GRMHD+GRFFE simulation data (right column) were averaged over the time range $t=7500 \,t_{\rm g}$ to $t=10000 \,t_{\rm g}$. The magenta contour indicates the $\sigma=1$ surface, and white contours indicate the azimuthal component of the vector potential $A_\phi$.
}
\label{fig:phiavg}
\end{figure*}

\begin{figure*}
\centering
\includegraphics[width=0.9\textwidth]{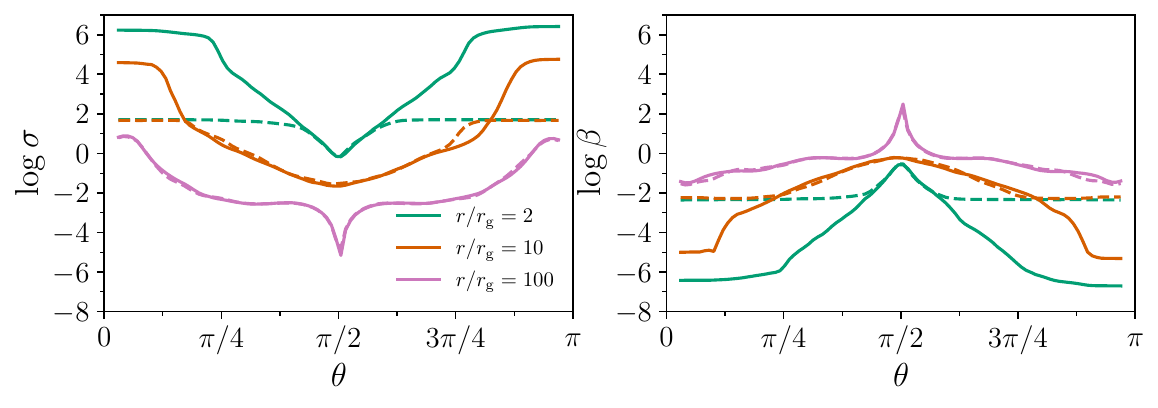}
\caption{
Polar profiles of the magnetization $\sigma$ (left) and ratio of thermal to magnetic pressure $\beta$ (right) at constant radii $r=2,10,100\,r_{\rm g}$ in both simulations. The simulation data were time and azimuth-averaged as in \autoref{fig:phiavg}. Profiles from the fiducial GRMHD simulation are displayed in dashed lines and profiles from the hybrid GRMHD+GRFFE simulation are displayed in solid color. }
\label{fig:phiavgrsli}
\end{figure*}

\begin{figure*}
\centering
\includegraphics[width=\textwidth]{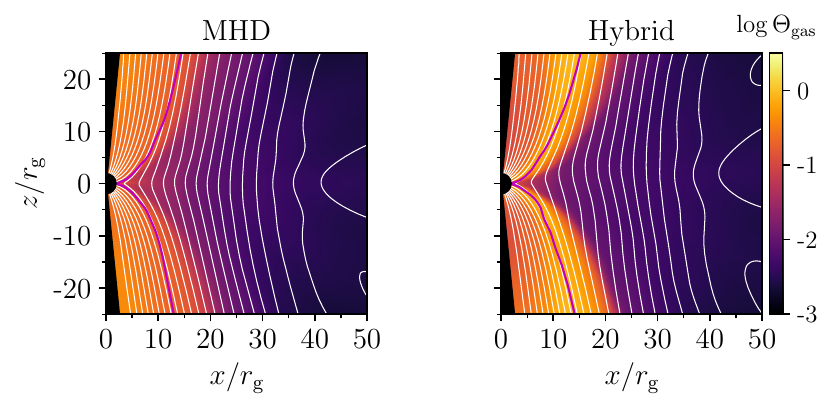}
\caption{
Time- and azimuth-averaged poloidal profiles of the dimensionless plasma temperature $\Theta_{\rm gas}$. The right plot shows averaged data from the standard GRMHD simulation and the right plot shows data from the hybrid GRMHD+GRFFE simulation. As in \autoref{fig:phisli}, The magenta contour in both panels indicates the $\sigma=1$ surface, and white contours indicate the azimuthal component of the vector potential $A_\phi$. 
}
\label{fig:phiavgtemp}
\end{figure*}

\begin{figure*}
\centering
\includegraphics[width=0.8\textwidth]{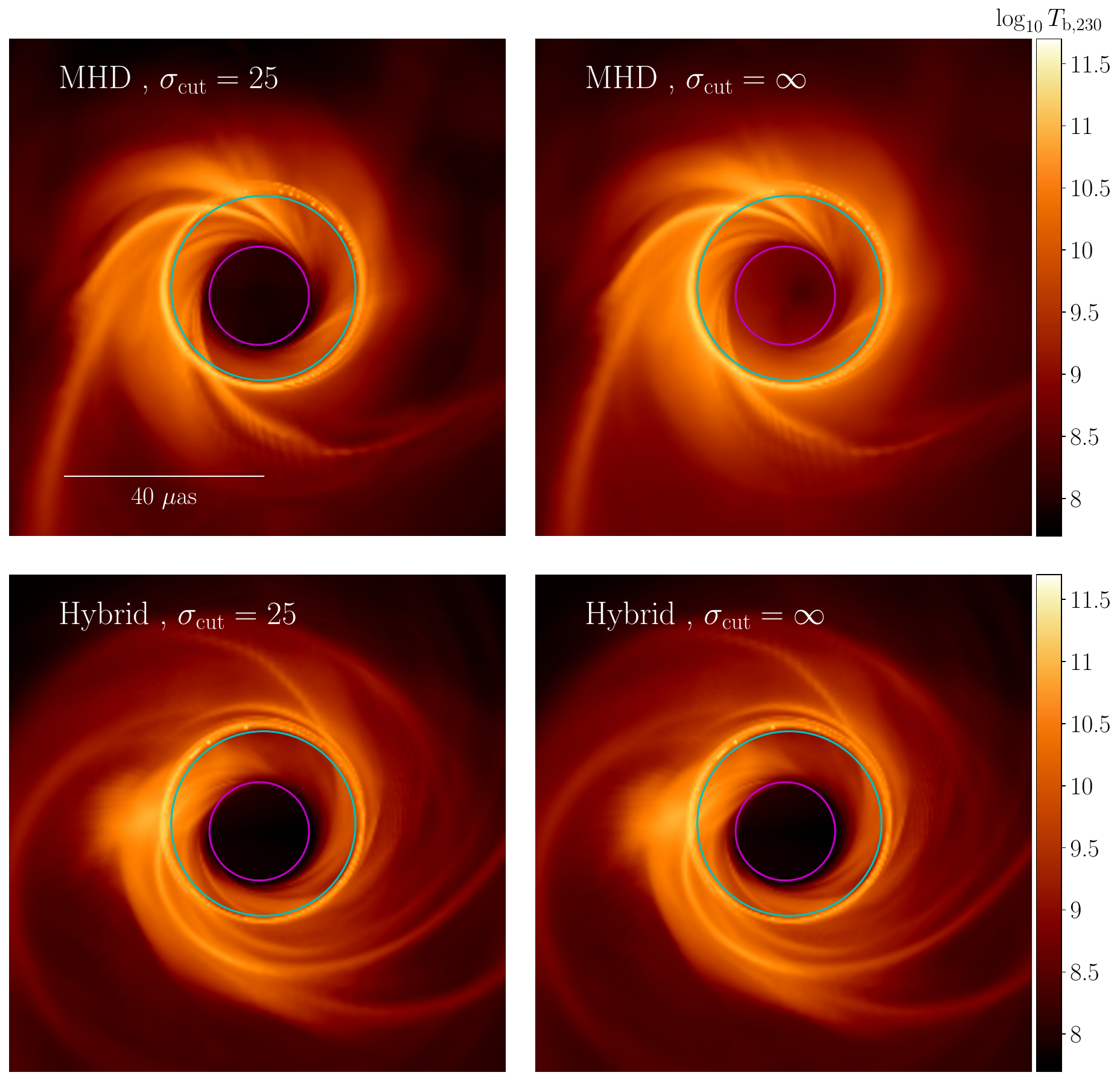}
\caption{
230 GHz images from both simulations with parameters (black hole mass, total flux density) scaled to match EHT observations of M87* \citep{PaperI}. The top column shows images from the standard GRMHD simulation taken at 
$t=8880\,t_{\rm g}$; the left column shows the image raytraced zeroing out emissivities wherever $\sigma>\sigma_{\rm cut}=25$, while in the right column no regions have their emissivities set to zero in the radiative transfer.  The bottom row shows corresponding images from a snapshot of the Hybrid GRMHD+GRFFE simulation at $t=8550\,t_{\rm g}$. In all images, the cyan contour indicates the critical curve or ``black hole shadow'' \citep{Bardeen1973a}; the magenta curve indicates the image of the equatorial event horizon, or ``inner shadow'' feature \citep{Chael2021}. All images are displayed in a log color scale.
}
\label{fig:images}
\end{figure*}

\begin{figure*}
\centering
\includegraphics[width=\textwidth]{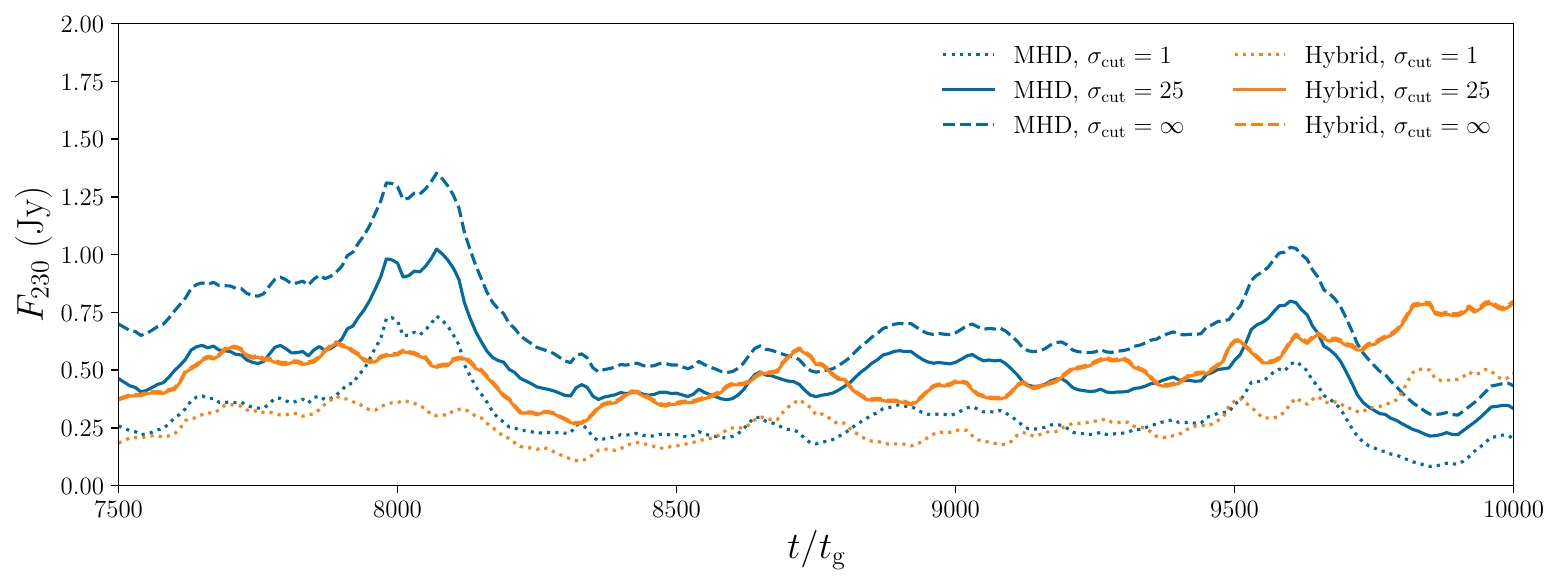}
\caption{
230 GHz simulation lightcurves over the range $7500 \,t_{\rm g}< t<10000\,t_{\rm g}$ with black hole mass mass and accretion rate scaled to match 2017 EHT observations of M87*. The blue lines represent lightcurves from images raytraced from the standard GRMHD simulation; the orange lines indicate data from the Hybrid GRMHD+GRFFE simulation. Dotted lines indicate that the lightcurves from images generated with $\sigma_{\rm cut}=1$; solid lines indicate the lightcurves from images generated with $\sigma_{\rm cut}=25$; dashed lines indicate lightcurves from images generated with no $\sigma_{\rm cut}$ applied.
}
\label{fig:lightcurves}
\end{figure*}

To test our new method for hybrid GRMHD+GRFFE simulations on a real black hole accretion problem, we performed two 3D simulations of magnetically arrested (MAD; \citealt{Bisnovatyi1976, NarayanMAD}) accretion discs around black holes with dimensionless spin $a_*=0.5$.

One simulation ("MHD") was performed with the standard GRMHD approach in \texttt{KORAL}, with a ceiling on the magnetization $\sigma_{\rm max}=50$ imposed in the normal observer frame \citep{McKinney2012}. The other simulation ("Hybrid") was performed with our new hybrid approach with a transition between GRMHD and GRFFE evolution at $\sigma_{\rm trans}=50$ and with transition width $w=0.1$. The setup of the other numerical floors and simulation parameters were otherwise identical in both simulations. In Appendix \ref{app:sigtranscompare}, we examine the effect of choosing higher and lower values of $\sigma_{\rm trans}$ in hybrid simulations with the same parameters.

\subsection{Simulation setup}
The parameters of the both simulations are reported in \autoref{tab:model_summary}. Both simulations were conducted on a uniform grid in modified Kerr-Schild coordinates similar to those used in the monopole test (\autoref{eq:mks}). The inner edge of the simulation grid is at $r=1.5 \,r_{\rm g}$ inside the event horizon $r_+\approx1.87\,r_{\rm g}$; the outer edge of the grid is at $r=1000_{\rm g}$. The resolution of the simulations were $160\times128\times98$ in radius, polar angle, and azimuth. 
Both simulations assume a gas adiabatic index $\adi=13/9$.

We initialized both simulations with a thick torus of plasma in hydrodynamic equilibrium following \citet{FishMonc}. The torus inner edge is at $r_{\rm in}=20 \,r_{\rm g}$ and the maximum pressure is at $r_{p,\mathrm{max}}=42\,r_{\rm g}$. The maximum density is set to $\rho_{\rm max}=1$ in code units; as the GRMHD equations are scale free, this normalization has no effect on the evolution.  The initial torus is threaded with a single loop of poloidal magnetic field from the vector potential \citep{Narayan2022}: 
\begin{align}
A_\phi &= \mathrm{Max}\left\{0\,,\, \left(\frac{\rho}{\rho_{\rm max}}\right)\left(\frac{r\sin\theta}{r_{\rm in}}\right)^3e^{-r/r_{\rm mag}}-0.2\right\}
\end{align}
where $r_{\rm mag}=400 \,r_{\rm g}$. The magnetic field in the initial loop is scaled such that the ratio of the maximum thermal pressure to maximum magnetic pressure in the disc is $\beta_{\rm max}=100$.
To initiate accretion, we seed the initial torus with Gaussian perturbations in the pressure with a fractional standard deviation of $2$ percent relative to the equilibrium \citet{FishMonc} value. 

Both simulations used second order PPM spatial reconstruction and LLF fluxes. Outflowing boundary conditions were used at the inner and outer radii, and reflecting boundary conditions are imposed at the polar axes. To control numerical instability from material reflecting off of the polar axes, we smoothly interpolate the poloidal velocity $u^\theta$ to zero across the two cells closest to the axis.

\subsection{Simulation results}

We ran both simulations to a final time $t_{\rm final}=10000 \,t_{\rm g}$ and saved snapshot files every $10 \,t_{\rm g}$. In \autoref{fig:scalars}, we show the evolution of the mass accretion rate and magnetic flux on the horizon from the beginning to the end of the simulation. At every time $t$ we
we calculate the mass accretion rate $\dot{M}$ and magnetic flux through the horizon  $\Phi_B$ by integrating over the horizon at $r=r_+$:
\begin{align}
\dot{M} &= -\int_\theta \int_\phi \rho u^r \sqrt{-g} \, d\theta d\phi \\
\Phi_B &= \int_\theta \int_\phi B^r \sqrt{-g}\, d\theta d\phi.
\end{align}
Note that we use $B^r = \Fd^{rt} = \B^r/\alpha$ in calculating the magnetic flux. Given these quantities, the dimensionless magnetic flux or "MAD parameter" is
\begin{equation}
\label{eq:madpar}
\phi = \frac{1}{2}\sqrt{\frac{4\pi\Phi^2_B}{\dot{M}}}.
\end{equation}
We use the factor $\sqrt{4\pi}$ to convert the magnetic field back to Gaussian units in computing $\phi$. Under this convention, MAD accretion discs tend to saturate at a normalized magnetic flux $\phi\approx50$ \citep{Tchekhovskoy11}.

\autoref{fig:scalars} shows that both the standard GRMHD and hybrid simulations take approximately $5000 \,t_{\rm g}$ from the start of the simulation to reach a steady-state in both $\dot{M}$ and $\phi$. In \autoref{tab:model_summary}, we report the averaged values of $\dot{M}$ and $\phi$ for both simulations over the second half of both simulations $5000\leq t/t_{\rm g}\leq 10000$. Both simulations saturate with an average value of $\phi\approx55$ and differ by $\approx7\%$.

Notably, when the magnetic field is first building up on the black hole horizon around $t=5000 \,t_{\rm g}$, the two simulations show somewhat different behavior. The hybrid simulation climbs to higher values of the magnetization $\phi$ up to $\phi\approx100$ before its first flux eruption event, and then settles down to steady state for the remainder of the simulation.  In a future work we will investigate higher resolution hybrid simulations and run them to longer times to see if larger values $\phi$ in the initial phase of MAD simulations is a generic feature of our hybrid method.

In \autoref{fig:phisli}, we show snapshot data in the poloidal plane from the two simulations of three quantities: the density $\rho$ (in dimensionless code units), the magnetization $\sigma$, and the ratio of the thermal to magnetic pressure $\beta$, defined as
\begin{equation}
\label{eq:beta}
\beta=\frac{2p}{b^2}.
\end{equation}
In \autoref{fig:phiavg} we show the same quantities from simulation data averaged in azimuth and over the time interval $7500\leq t/t_{\rm g}\leq10000$. In plotting the averaged profiles of $\sigma$ and $\rho$, we average the magnetic pressure, density, and thermal pressure independently and then take the appropriate ratios in \autoref{eq:sigma} and \autoref{eq:beta}.

Both \autoref{fig:phisli} and \autoref{fig:phiavg} show that the global structure of the two simulations within $r\lsim50_{\rm g}$ is substantially similar, particularly for low $\sigma$ regions in the bulk of the accretion disc and corona region. Inside the magnetized jet, the density in the hybrid simulation is allowed to fall to much lower values and the magnetization correspondingly rises to a maximum $\sigma\approx10^6$ close to the black hole; in the standard GRMHD simulation, the magnetization $\sigma\lsim50$ everywhere because our ceiling limits the magnetization below this value.\footnote{Because the ceiling on $\sigma$ is imposed in the normal observer frame, the fluid frame $\sigma$ reported in these figures may exceed $\sigma_{\rm max}=50$} 
In addition to lowering the overall value of $\sigma$ in the jet region, the standard GRMHD simulation also reaches much lower levels of $\beta$ in this region and develops a signature halo of high density $\rho$ around the black hole purely from floor material (upper left panel of \autoref{fig:phiavg}). In \autoref{fig:phiavgrsli}, we directly compare $\sigma$ and $\beta$ on different contours of constant radius $r$ in the averaged data from the two simulations. 
The average values of $\sigma$ and $\beta$ in the simulations match well at $r=100 \,r_{\rm g}$ and around the equatorial plane at smaller radii $r=2\,r_{\rm g}$ and $r=10r_{\rm g}$; at the smaller radii close to the poles, the GRMHD simulation values of $\sigma$ and $\beta$ level off at around $50$ and $0.01$, respectively. 
By contrast, in the hybrid simulation, the average $\sigma$ can rise to $10^{6}$ near the black hole, with $\beta$ falling correspondingly to $\approx10^{-6}$. 

In \autoref{fig:phiavgtemp} we show time and azimuthal average profiles of the dimensionless plasma temperature $\Theta_{\rm gas}$ from both simulations. GRMHD simulations typically show very high temperatures $\Theta_{\rm gas}\approx 1$ in the jet region, which we also see in our comparison GRMHD run. The core of the jet in our hybrid simulation is kept cooler by our imposition of adiabatic fluid evolution in the force-free region of the simulation; $\Theta_{\rm gas}\approx 0.1$. 
The exact profile of temperature in the transition region near $\sigma=1$ between the jet and disc can be important in determining observable image features and spectra from synchrotron emission \citep[e.g.][]{Chael19,PaperV}. It is still not entirely clear if the rapid increase in temperature with $\sigma$ in this region is entirely physical, or if the rapid decrease in plasma density causes issues in the numerical treatment of dissipation \citep{Ressler17}.
As in standard GRMHD, our hybrid method shows a significant increase in the plasma temperature as $\sigma$ increases, with temperatures near the $\sigma=1$ surface slightly exceeding the values seen in standard GRMHD. The jet core is slightly cooler in the hybrid simulation than in the standard GRMHD approach. 

\subsection{230 GHz images and lightcurves}
\label{sec:images}

We next consider the implications of our proposed method for hybrid GRMHD+GRFFE simulations for simulated observables of black hole accretion flows. We produce simulated 230 GHZ images of M87* from both of our simulations using the GR radiative transfer code \texttt{ipole} \citep{ipole} over the time period $7500 \,t_{\rm g}<t<10000 \,t_{\rm g}$. In producing these images set the temperature of the emitting electrons using the standard \citet{Mosc16} $R_{\rm high}$-$R_{\rm low}$ prescription. The ion-to-electron temperature ratio $R$ is taken as a function of the local $\beta$:
\begin{equation}
R = R_{\rm high}\frac{\beta^2}{1+\beta^2} + R_{\rm low}\frac{1}{1+\beta^2}.
\end{equation}
We fix $R_{\rm low}=1$, $R_{\rm high}=20$ in the images shown here. Given $R$, the number density $n_e$ and temperature $T_e$ of the emitting electrons are calculated from the simulation mass density $\rho$ and internal energy $\ugas$ as   
\begin{align}
n_e &= n_i = \rho/m_p \\
T_e &= \frac{u_{\rm int}}{n_e k_{\rm B}}\left(\frac{1}{\Gamma_e-1} + \frac{R}{\Gamma_i-1}\right)^{-1}.
\end{align}
We assume the adiabatic index of electrons is $\Gamma_e=4/3$ and the adiabatic index of the ions is $\Gamma_i=5/3$. Given the electron number density and temperature, as well as the magnetic field strength and orientation, \texttt{ipole} solves the radiative transfer equation for synchrotron emission along the curved geodesic trajectories of light around our $a_*=0.5$ Kerr black hole to produce a simulated image. 

Because GRMHD simulations are scale-free, we scale the black hole mass and accretion rate in our simulations to values appropriate for M87* \citep{PaperI}. We take $M=6.5\times10^{9}M_\odot$ and set the distance to M87* as $D=16.8$ Mpc \citep{PaperVI}. 
We fix the inclination angle of the spin axis to $\theta_o = 163\deg$ \citep{Walker2018}.
We then  scale the mass accretion rate independently in each simulation so that the median flux density at 230 GHz over all of the images we raytrace is 0.5 Jy. The field of view of our images is $160\mu$as ($\approx43 \,r_{\rm g}$), and we use a pixel size of $0.5\mu$as. 

Synchrotron images from GRMHD simulations typically do not include contributions to the emission or absorption from plasma with a magnetization $\sigma>\sigma_{\rm cut}$, where typically $\sigma_{\rm cut}=1$ \citep[e.g.][]{Prather2023}. The value of $\sigma_{\rm cut}$ is typically set lower than the simulation ceiling value $\sigma_{\rm max}\approx50-100$ for two reasons. First, we wish to avoid contributions to the simulated image from regions where the plasma density is dominated by floor material. Second, it is frequently a concern that the plasma properties from GRMHD simulations may not be reliable even in regions of intermediate magnetization $1<\sigma<\sigma_{\rm max}$ where the simulation does not require floors for numerical stability. In this work we follow \citet{Chael19} in setting a fiducial $\sigma_{\rm cut}=25$ instead of $\sigma_{\rm cut}=1$. We find that in both simulations there is significant emission from intermediate magnetization regions $1<\sigma<25$ along the jet sheath, but there is no significant emission from the jet core at $\sigma>25$ in the hybrid simulation.

For both the GRMHD and hybrid simulation, we produce images with $\sigma_{\rm cut}=1$,  $\sigma_{\rm cut}=25$, and $\sigma_{\rm cut}=\infty$, where in the latter case we include synchrotron emission and absorption from all material in the simulation domain. In \autoref{fig:images} we show a comparison between image snapshots from both simulations at all three values of $\sigma_{\rm cut}$. It is evident that the hybrid simulation shows no change in image structure between $\sigma_{\rm cut}=25$ and $\sigma_{\rm cut}=\infty$; by contrast, in the MHD simulation the $\sigma_{\rm cut}$ shows a significant contribution ($\sim 20\%$ of the total flux density) from the floor material in the forward jet, forming a haze in front of the black hole shadow and inner shadow \citep{Chael2021} features. In \autoref{fig:lightcurves} we show simulation lightcurves over the range $7500 \,t_{\rm g}<t<10000 \,t_{\rm g}$ from both simulations at both values of $\sigma_{\rm cut}$. For the hybrid simulation, the lightcurves for both $\sigma_{\rm cut}=25$ and $\sigma_{\rm cut}=\infty$ are nearly identical, while the overall flux density of the GRMHD lightcurve is increased with the addition of floor material when increasing $\sigma_{\rm cut}$ from $25$ to $\infty$. In addition to allowing GRMHD simulations to stably evolve Poynting dominated jets to high values of $\sigma$, our proposed hybrid GRMHD+GRFFE simulation method may remove a potential source of systematic error in producing simulated images from GRMHD simulation by removing the requirement to set a $\sigma_{\rm cut}$ value when performing radiative transfer.

\section{Discussion and Conclusions}
\label{sec:discussion}

In this paper we have presented a proposed approach for stably evolving GRMHD simulations of black hole accretion to high values of the magnetization $\sigma$ by switching to an augmented set of force-free equations in this region. Our method evolves the force free equation in the finite volume formulation of \citet{McKinney06}; it solves for the field-parallel velocity, plasma density, and energy density in the magnetically dominated region with an approximate set of decoupled equations that do not back-react on the electromagnetic field evolution. We propose a method for joining force-free evolution to standard GRMHD evolution at a specified transition $\sigma_{\rm trans},$ either via a sharp transition or a smooth average of the recovered GRMHD primitives across a certain range of $\sigma$ values. 

We tested our new hybrid GRMHD+GRFFE code on a set of test problems and showed that it can accurately evolve a highly magnetized plasma across the interface $\sigma_{\rm trans}$; in the Bondi test problem in particular (\autoref{sec:bondi}), we see a smaller error in the plasma density, energy density, and velocity evolved under our hybrid scheme than in a standard GRMHD approach. We then compared our hybrid method with a standard GRMHD set-up in a magnetically arrested simulation of an accreting black hole with the same initial conditions and resolution. Our hybrid method produced a MAD disc with average properties similar to standard GRMHD in low magnetization regions, but the absence of hard ceilings on $\sigma$ allows the jet funnel to evacuate and the magnetization to reach $\sigma\approx 10^6$. Unlike the GRMHD simulations which are contaminated by floor material in high-magnetization regions, the hybrid simulations produce simulated submillimeter synchrotron images that are insensitive to the $\sigma_{\rm cut}$ parameter above $\sigma_{\rm cut}\approx25$. In other words, our hybrid method naturally produces an evacuated force-free jet funnel in GRMHD simulations that does not contribute to the synchrotron emission simulated for comparison to EHT images. 

Our proposed hybrid method is a relatively straightforward addition to the framework of finite-volume GRMHD codes. Because the conserved-to-primitive inversion $\mathbf{P}_{\rm FFE}(\mathbf{U})$ for the force-free quantities is analytic in the cold approximation that we use, and because we only perform the additional GRFFE conserved to primitive inversion step in a small part of the simulation domain, our hybrid scheme does not significantly increase the computational expense of standard GRMHD at the resolution we considered for our 3D MAD simulations. The primary additional computational cost comes from computing fluxes for the additional $\mathbf{P}_{\rm FFE}$ quantities at each cell walls. In our current implementation we compute these $\mathbf{F}(\mathbf{P}_{\rm FFE})$ everywhere, but in the future we plan to increase the simulation efficiency further by limiting the domain where we compute $\mathbf{F}(\mathbf{P}_{\rm FFE})$ only to regions where they are required in a given time step. 

Our method is not the first to consider matching GRMHD to GRFFE in high magnetization regions in a single simulation volume. Notably, \citet{Paschalidis13} considered hybrid GRMHD and GRFFE simulations of neutron star magnetospheres. They use a similar finite-volume approach motivated by \citet{McKinney06} to the method presented in this paper to evolve the force-free region. However, they do not evolve the field-parallel velocity or fluid quantities in the force free region, and they fix the transition between force-free and GRMHD evolution to the surface of the neutron star. By contrast, in our proposed method the boundary between the GRMHD and GRFFE simulation regions evolves with the simulation and propagating jet from the black hole; plasma is allowed to flow in between the two regions by solving the parallel velocity and continuity equations. 

Later, \citet{Parfrey2017,Parfrey2023} conducted 2D and 3D hybrid neutron star magnetosphere simulations using a GRMHD code adapted for force-free evolution in the high magnetization region. In the simulations of \citet{Parfrey2017,Parfrey2023},
the GRMHD equations are evolved everywhere, but the code damps the field-parallel velocity in the magnetosphere region. The region where the code transitions to force-free behavior is determined by a combination of an advected passive scalar distinguishing the initial accretion flow and magnetosphere and a function of the coordinates that goes to zero outside the light cylinder. Recently, \citet{Phillips2023} presented a novel operator splitting method in special relativistic MHD that solves for the MHD evolution of the velocity and magnetic field over a timestep as a correction to the value obtained from force-free evolution. This method can stably evolve SRMHD problems to high magnetizations without requiring a pre-defined transition $\sigma_{\rm trans}$ between MHD and FFE regions, but it has not yet been applied to 3D GRMHD simulations.

By coupling GRMHD to GRFFE, the method proposed in this paper can stably evolve jets in GRMHD simulations of black hole accretion to a magnetization four orders of magnitude larger than the largest $\sigma$ achieved in standard GRMHD. Other techniques or improvements to the method presented here may further increase the reliability of GRMHD in the jet region and open up a range of new questions to investigation using GRMHD simulations, including; do evacuated jets show more pronounced limb brightening \citep[e.g.][]{Lu2023} in simulated VLBI images than standard GRMHD images \citep{Chael19, Fromm2021}? Do GRMHD jets evolved without density floors reach higher Lorentz factors or have significantly different shapes when evolved for long times, and how do they compare to observation of M87* \citep{Nakamura2018,Park2019}? Can we add self-consistent pair-production \citep[e.g.][]{Broderick15,Wong2021} to simulations of GRMHD jets to fill in the core and investigate the importance of pairs to EHT images? As GRMHD simulations of black hole accretion become more sophisticated, increasing in resolution and adding additional radiative and thermodynamic physics, it is essential to revisit the standard approach to evolving the MHD equations in magnetically dominated regions.

\section*{Acknowledgements}
We thank the anonymous referee for their helpful comments and suggestions, which significantly improved the paper. We thank George Wong and Eliot Quataert for useful conversations, and Angelo Ricarte and Richard Qiu for providing the Python code used to produce the 230 GHz images in \autoref{sec:images}. This work used Stampede2 at TACC through allocation TG-AST190053 from the Extreme Science and Engineering Discovery Environment (XSEDE), which was supported by National Science Foundation grant number \#1548562. Computations in this paper were run on the FASRC cluster supported by the FAS Division of Science Research Computing Group at Harvard University

\section*{Data Availability}
The data from the simulations reported here is available on request. The \texttt{KORAL} code is available at \url{https://github.com/achael/koral_lite}.



\bibliographystyle{mnras}
\bibliography{allrefs} 


\appendix

\section{Degenerate Electromagnetic fields}
\label{app:fields}

For arbitrary electromagnetic fields, the full Faraday tensor and Maxwell tensors can be decomposed into the normal observer electric and magnetic fields as
\begin{align}
    F^{\alpha\beta}&=\eta^\alpha \E^\beta - \eta^\beta E^\alpha - \epsilon^{\alpha \beta\gamma\delta} \B_\gamma \eta_\delta , \nonumber \\
    \Fd^{\alpha\beta}&=-\eta^\alpha \B^\beta + \eta^\beta \B^\alpha -  \epsilon^{\alpha \beta\gamma\delta} \E_\gamma \eta_\delta.
    \label{eq::Faradays1}
\end{align}
The contractions of $F$ and $\Fd$ are invariant scalars: 
\begin{align}
    F_{\mu\nu}\Fd^{\mu\nu} &= 4\E^\mu \B_\mu \\
    F_{\mu\nu}F^{\mu\nu} &= 2(\B^2-\E^2). 
\end{align}
The general form of an electromagnetic stress-energy tensor is 
\begin{align}
    T^{\mu\nu}_\mathrm{EM} &= g_{\alpha\beta}F^{\mu \alpha} F^{\nu \beta} - \frac{1}{4}g^{\mu \nu}F^{\alpha\beta}F_{\alpha \beta},
    \label{eq:Tem}
\end{align}
In terms of the normal observer frame fields $\E^\mu$ and $\B^\mu$, a general $T^{\mu\nu}_\mathrm{EM}$ can be expressed as
\begin{align}
    T^{\mu\nu}_\mathrm{EM} &= \left(\B^2 + \E^2\right)\left(\eta^\mu\eta^\nu + \frac{1}{2}g^{\mu\nu}\right) \nonumber \\
    &- \left(\B^\mu\B^\nu + \E^\mu \E^\nu\right) \nonumber \\
    &-\eta_\alpha \E_\beta \B_\kappa \left(\eta^\mu \epsilon^{\nu\alpha\beta\kappa} + \eta^\nu \epsilon^{\mu\alpha\beta\kappa}\right).
    \label{eq:TemEB}
\end{align}
The conditions for degeneracy and magnetic domination are
\begin{align}
    \Fd_{\mu\nu}F^{\mu\nu}&=0&&\text{(degenerate)},\\
    F^{\mu\nu}F_{\mu\nu} &>0&&\text{(magnetically dominated)}.
    \label{eq:Conditions}
\end{align}    
A degenerate EM field has an infinite family of timelike vectors in the kernel of $F^{\mu\nu}$ such that the electric field in these frames $e^\mu=u_\nu F^{\mu\nu}=0$.
Degeneracy implies that the normal observer electric and magnetic fields are orthogonal, $\E_\mu\B^\mu=0$, and magnetic domination implies that the fluid frame magnetic energy density is always positive,  $u_\mu u_\nu T^{\mu\nu} = \frac{1}{2}b^2 >0$. 

In GRMHD, the fluid frame magnetic field is given by \autoref{eq:bmu}, which can be expressed in component form as  
\begin{align}
 b^0 &= \frac{1}{\alpha}u_i\B^i = u_i B^i, \nonumber \\
 b^i &= \frac{1}{\gamma}\left(\B^i + \alpha b^0 u^i\right) = \frac{1}{u^0}\left(B^i + b^0 u^i\right),
\end{align}
where $B^i=\Fd^{i0}=\B^i/\alpha$ is the lab frame magnetic field that is often used in GRMHD codes \citep{Gammie03} instead of the normal observer frame fields. The time component of the fluid-frame magnetic field $b^0$ depends only on the part of the normal-observer frame field that is parallel to the fluid velocity.

\section{Field-parallel velocities}
\label{app:velpar}

For a degenerate, magnetically dominated field in either GRMHD and GRFFE, the drift velocity $u^\mu_\perp$ for a given coordinate system is the unique frame where the electric field vanishes and the Lorentz factor is minimized, as the motion is entirely perpendicular to the magnetic field in that frame. In terms of the normal observer frame three-velocity \autoref{eq:normalobsvel}, the drift frame is \citep{McKinney06}:
\begin{equation}
    \vperp^i = \frac{\alpha}{\sqrt{-g}\B^2}[ijk]\E_j\B_k.
    \label{eq:uMcKinney}
\end{equation}
\autoref{eq:uMcKinney} is the relativistic analogue of the non-relativistic drift velocity $(\mathbf{E}\times\mathbf{B}) / \mathbf{\B}^2$. Note that in \autoref{eq:uMcKinney} the magnitude $\vperp^2=\E^2/\B^2$. In order for the drift frame to be timelike, $\vperp^2<1$, the field must be magnetically dominated. 

For degenerate fields, neither $\Fd^{\mu\nu}$ nor $T^{\mu\nu}$ depend on the part of the fluid velocity parallel to the magnetic field. Thus, the force-free equations of motion do not uniquely determine $u^\mu$, as we can add arbitrary components of the velocity along the field-line and still end up with the same $\Fd^{\mu\nu}$ and $T^{\mu\nu}$ that solve the force-free equations.  We define field parallel and field-perpendicular three-velocities as 
\begin{subequations}
\begin{align}
    \vpar^\mu &= \vpar \frac{\B^\mu}{\sqrt{\B^2}}, \\
    \vperp^\mu &= \tilde{v}^\mu - \vpar^\mu.
    \label{eq:vparvperp}
\end{align}
\end{subequations}

The normal observer frame three-velocities $\vpar^\mu$, $\vperp^\mu$ are orthogonal and have corresponding Lorentz factors $\lpar$, $\lperp$ given by \autoref{eq:gammafromv}.\footnote{If $\upar^\mu=\lpar\vpar^\mu$ and $\uperp^\mu=\lperp\vperp^\mu$, $\tilde{u} = \left(\gamma/\lperp\right)\uperp^\mu + \left(\gamma/\lpar\right)\upar^\mu$.}
Because the velocities are orthogonal, $\tilde{v}^2 = \vpar^2 + \vperp^2$, and the total Lorentz factor is
\begin{align}
    \gamma^2 &= \left(1-\vpar^2-\vperp^2\right)^{-1} = \left(\lpar^{-2} + \lperp^{-2} -1\right)^{-1}.
\end{align}
Note that the time component of the fluid-frame magnetic field is
\begin{equation}
   b^0 = \frac{1}{\alpha}u_\mu \B^\mu = \frac{1}{\alpha}\tilde{u}_\mu \B^\mu = \frac{\gamma}{\alpha}\vpar\sqrt{\B^2}. 
\end{equation}
We can thus express the fluid-frame magnetic field in terms of the parallel velocity $\vpar$ as
\begin{equation}
    b^\mu = \frac{1}{\gamma}\B^\mu + \left(\vpar\sqrt{\B^2}\right)u^\mu.
\end{equation}
Then, fluid-frame squared magnetic field strength is
\begin{align}
    b^2 &= \B^2\left[\frac{1}{\gamma^2} + \vpar^2\right] = \frac{\B^2}{\lperp^2}.
\end{align}
The fluid-frame magnetic field energy density $b^2$ thus only depends on the perpendicular velocity.\footnote{We can also show if in the fluid frame $e^\mu=0$, then in the normal observer frame $\E^2 = \tilde{v}^2_\perp \B^2.$} 
Using \autoref{eq:TemDeg}, we can show that, in fact, the full stress-energy tensor $T^{\mu\nu}_\mathrm{EM}$ only depends on the perpendicular velocity: 
\begin{equation}
    T^{\mu\nu}_\mathrm{EM} = b^2u^\mu_\perp u^\nu_\perp + \frac{1}{2}b^2g^{\mu\nu} - b^\mu_\perp b^\nu_\perp,
\end{equation}
where
\begin{align}
    u^\mu_\perp &= \lperp \left(\vperp^\mu + \eta^\mu\right), \\
    b^\mu_\perp &= {\gamma_\perp}{\B^\mu}.
\end{align}
Thus, any parallel velocity $\vpar$ does not enter in to the degenerate electromagnetic stress-energy tensor. As a result, in force-free electrodynamics, the equations of motion do not determine the evolution of the field-parallel part of the fluid velocity.

\section{GRMHD Conserved to Primitive Inversion}
\label{app:u2p}
The standard way to perform the numerical inversion from conserved quantities $\mathbf{U}$ to primitive quantities $\mathbf{P}$ in GRMHD was described by \citet{Noble2006}. In this method, we rearrange the equations relating the conserved and primitive quantities so that we only need to numerically solve for one variable, the observer-normal frame enthalpy $W$ defined as
\begin{equation}
    W \equiv \gamma^2 h. 
\end{equation}
To do this, we first project the conserved energy flux $\mathcal{Q}_\mu=-\eta_\nu T^{\nu}_{\mu}$ to obtain $\tilde{\mathcal{Q}}^\mu$, the energy flux perpendicular to the normal observer (\citealt{Noble2006}, Equation 30):
\begin{equation}
    \tilde{\mathcal{Q}}^\mu = j^\mu_{\;\;\nu}Q^\nu =\left(W  + \B^2\right)\tilde{v}^\mu - \frac{\left(\B^\nu \tilde{Q}_\nu\right)\B^\mu}{W}.
    \label{eq:Qofv}
\end{equation}
Note that none of the components of $\tilde{\mathcal{Q}}^\mu$ depend on the time component, $\mathcal{Q}^0$.\footnote{In particular, $\tilde{\mathcal{Q}}^0 = 0$ and $\tilde{\mathcal{Q}}^i = g^{i\nu}\tilde{\mathcal{Q}}_\nu$, where $\mathcal{Q}_i = \tilde{\mathcal{Q}}_i$ and $\tilde{\mathcal{Q}}_0 = \beta^i \mathcal{Q}_i$.}
After projecting,  we can express the normal frame velocity magnitude $\tilde{v}^2$ in terms of  $\tilde{\mathcal{Q}}^2$ via (\citealt{Noble2006}, Equation 27):
\begin{equation}
    \tilde{\mathcal{Q}}^2 = \tilde{v}^2\left(\B^2 + W \right)^2 - \frac{\left(\B^\mu \tilde{\mathcal{Q}}_\mu\right)^2\left(\B^2 + 2W\right)}{W^2}.
    \label{eq:Q2equation}
\end{equation}
\autoref{eq:Q2equation} gives an analytic relationship for $\tilde{v}^2(W)$ which depends on the spatial components of the conserved momentum $\mathcal{Q}_i = \alpha T^0_i$. 

To solve for $W$ we need one more equation for the conserved energy. This is:
\begin{equation}
    \mathcal{U} \equiv -\eta_\mu \mathcal{Q}^\mu = \alpha^2 T^{00} = W - p + \frac{\B^2}{2}\left(1+\tilde{v}^2\right) + \frac{\left(\B^\mu \tilde{\mathcal{Q}}_\mu\right)^2}{2W^2}.
    \label{eq:Wequation}
\end{equation}
In standard GRMHD conserved-to-primitive inversion using the \citet{Noble2006} method, we numerically solve \autoref{eq:Wequation} for $W$, using \autoref{eq:Q2equation} to obtain $\tilde{v}^2(W)$. Once we have solved for $W$ and $\tilde{v}^2(W)$, we can compute the Lorentz factor $\gamma$ (\autoref{eq:gammafromv}), the enthalpy $h$ from $W=\gamma^2 h$, the density $\rho$ from the conserved quantity $\mathcal{D}=\gamma\rho$, and the energy density $\ugas$ from the equation of state $\adi$ and the definition of the enthalpy $h=\rho+u+p$. Finally, \autoref{eq:Qofv} is used to find the velocities $\tilde{v}^i$.

We can also write the conserved quantities $\tilde{Q}^i$ and $\tilde{U}$ in terms of the parallel and perpendicular velocities and the magnetic field.  First, note that
\begin{equation}
    \B^\mu \mathcal{Q}_\mu = h\gamma \B^\mu u_\mu = h \gamma^2 \vpar \sqrt{\B^2}.
    \label{eq:BdotQ}
\end{equation}
Then, plugging \autoref{eq:BdotQ} into \autoref{eq:Qofv} and \autoref{eq:Wequation}, we can show that
\begin{align}
\label{eq::conserved2}
    \tilde{\mathcal{Q}}^i &= W \left(\vpar^i + \vperp^i\right) + \B^2\vperp^i, \\
    \mathcal{U} &= W - p + \frac{\B^2}{2}\left(1 + \vperp^2\right).
    \label{eq:QandUv2}
\end{align}
Again, we see that the electromagnetic parts of the conserved quantities $\mathcal{Q}_i$ and $\mathcal{U}$ only depend on the \emph{perpendicular} parts of the velocity. 


\section{GRMHD parallel momentum equation}
\label{app:pmom}

In this section we derive the general equation for the conservation of stress-energy parallel to the magnetic field in GRMHD, \autoref{eq:bpargeneral}. This derivation is based on a similar derivation in the Appendix of \citet{Camenzind1986}.

First, we contract the fluid velocity $u^\alpha$ with the homogeneous Maxwell equation for GRMHD (\autoref{eq:MaxHom}):
\begin{align}
0 &= u_\alpha \nabla_\beta \Fd^{\alpha\beta} \nonumber \\
&= u_\alpha \nabla_\beta \left(b^\alpha u^\beta - b^\beta u^\alpha\right) \nonumber\\
&= u^\alpha u^\beta \nabla_\beta b_\alpha  + \nabla_\beta b^\beta \nonumber\\
&= -b_\alpha u^\beta \nabla_\beta u^\alpha + \nabla_\alpha b^\alpha
\end{align}
where we used the fact that $u^\alpha b_\alpha=0$ and $u^\alpha u_\alpha = -1$. Therefore, for degenerate magnetically dominated fields
\begin{equation}
    \nabla_\alpha b^\alpha = a_\alpha b^\alpha, 
    \label{eq:divb}
\end{equation}
where the acceleration along the velocity direction is $a^\alpha \equiv u^\beta\nabla_\beta u^\alpha.$

Next, we contract $b^\alpha$ with the conservation of stress-energy equation  (\autoref{eq:GRMHDT}). We find first that contracting $b_\alpha$ with the divergence of $T^{\alpha\beta}_{\rm EM}$ gives zero  in all cases:
\begin{align}
b_\alpha \nabla_\beta T^{\alpha\beta}_{\rm EM} &= \frac{1}{2}b^\alpha \nabla_\alpha b^2 + b_\alpha \nabla_\beta \left(b^2 u^\alpha u^\beta\right) - b_\alpha \nabla_\beta \left(b^\alpha b^\beta\right)\nonumber\\
&= \frac{1}{2}b^\alpha \nabla_\alpha b^2 + b^2 b_\alpha a^\alpha - b_\alpha \nabla_\beta \left(b^\alpha b^\beta\right)\nonumber\\
&= \frac{1}{2}b^\alpha \nabla_\alpha b^2  - b^\alpha b^\beta \nabla_\beta b_\alpha \nonumber \\
& = 0.
\end{align}
In the second line we again used the fact that $b^\alpha u_\alpha=0$ and in the third line we used the relation for the divergence of $b^\alpha$ from Maxwell's equations, \autoref{eq:divb}. 

Next we contract $b^\alpha$ with the divergence of $T^{\alpha\beta}_{\rm fluid}$:
\begin{align}
b_\alpha \nabla_\beta T^{\alpha\beta}_{\rm fluid} &= b_\alpha \nabla_\beta \left(hu^\alpha u^\beta\right) + b^\beta \nabla_\beta p \nonumber\\
&= h b_\alpha a^\alpha + b^\beta \nabla_\beta p,
\end{align}

where we again used the fact that $u^\alpha$ is orthogonal to $b^\alpha$, and the definition of the acceleration $a^\alpha$. So, finally, from the conservation of total stress-energy and the relation \autoref{eq:divb}:
\begin{align}
    0 &= b_\alpha \nabla_\beta \left(T^{\alpha\beta}_{\rm EM} + T^{\alpha \beta}_{\rm fluid}\right) \nonumber \\
      &= b^\alpha\nabla_\alpha p + h \,\nabla_\alpha b^\alpha,
      \label{eq:bpargeneral2}
\end{align}
which gives the GRMHD relation for the field-parallel velocity used in the main text, \autoref{eq:bpargeneral}. 

To obtain the adiabatic limit of \autoref{eq:bpargeneral2}, we use the thermodynamic relation: 
\begin{align}
    \partial_\alpha\left(\frac{h}{\rho}\right) = \left(\frac{p}{\rho}\right) \partial_\alpha s + \left(\frac{1}{\rho}\right)\partial_\alpha p, 
    \label{eq:thermorelation}    
\end{align}
which we can easily verify from the definition of the entropy, \autoref{eq:entropydef}. If we assume the evolution of the gas is adiabatic and the entropy-per-particle is constant along field lines then $b^\alpha \nabla_\alpha s = 0$. We can then use the relation \autoref{eq:thermorelation} to replace the $\nabla_\alpha p$ term in \autoref{eq:bpargeneral}, and we find that
\begin{equation}
        \nabla_\alpha \mu b^\alpha=0,
\end{equation} 
which is \autoref{eq:vparadi}.

To take the cold limit of \autoref{eq:bpargeneral2}, we simply assume $\left| b^\alpha\nabla_\alpha p\right| << \left|h \,\nabla_\alpha b^\alpha,\right|$. In this case,
\begin{equation}
\nabla_\alpha b^\alpha=0,
\end{equation}
which is the form of the equation for the parallel velocity we use in the force-free region for the MAD simulations reported in this paper. 

\begin{figure*}
\centering
\includegraphics[width=.9\textwidth]{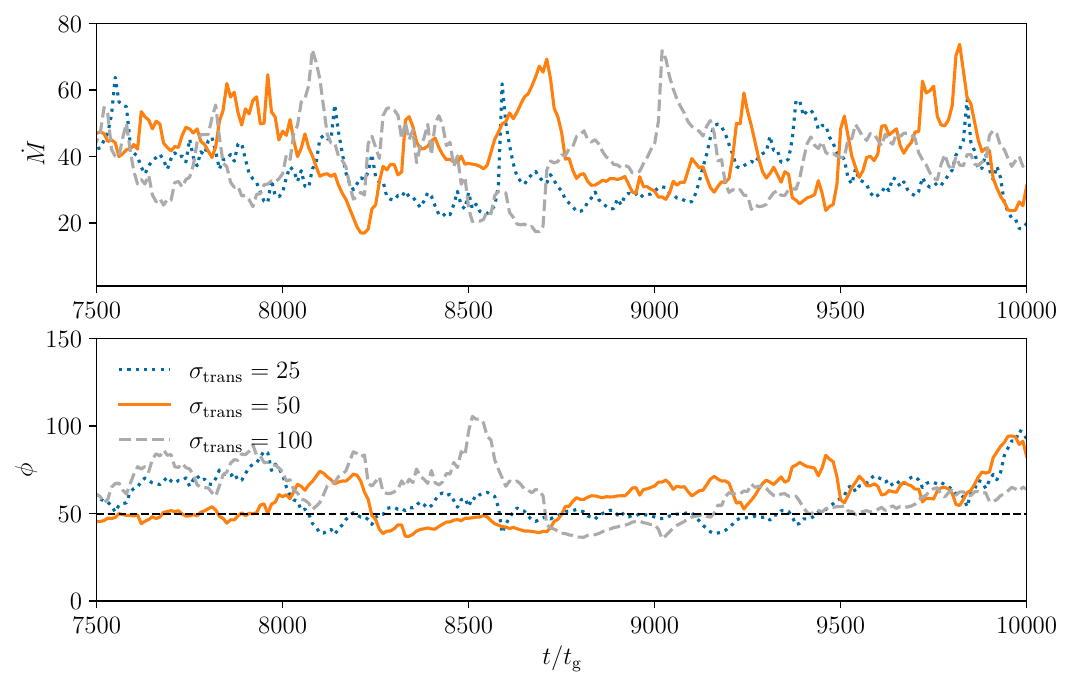}
\caption{
Evolution of the mass accretion rate $\dot{M}$ (top) and dimensionless magnetic flux through the horizon $\phi$ (bottom; \autoref{eq:madpar}) in the hybrid simulations with $\sigma_{\rm trans}\in \left\{25,50,100\right\}$ from $t=7500 \, t_{\rm g}$ to $t=10000 \, t_{\rm g}$.
}
\label{fig:scalarsapp}
\end{figure*}

\begin{figure*}
\centering
\includegraphics[width=0.9\textwidth]{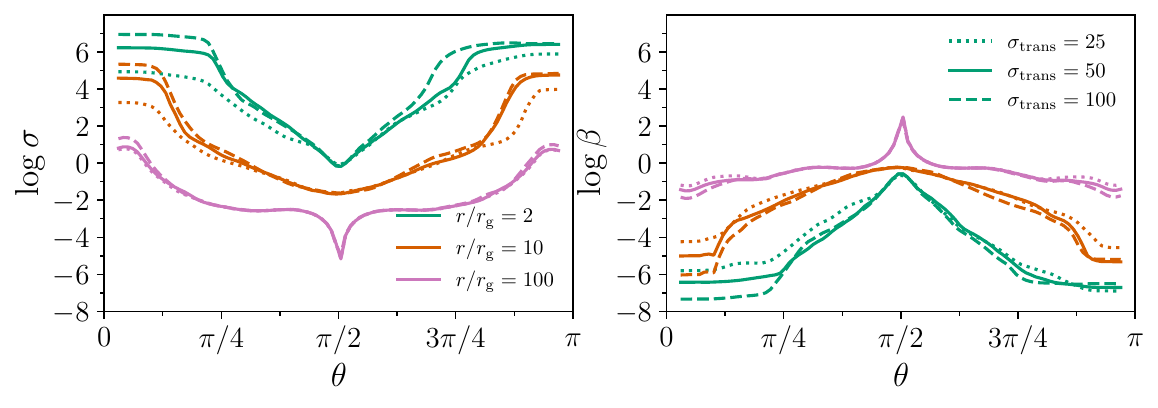}
\caption{
Polar profiles of the magnetization $\sigma$ (left) and plasma-$\beta$ (right) at constant radii $r=2,10,100\,r_{\rm g}$ in three hybrid simulations with $\sigma_{\rm trans}=25$ (dotted), $\sigma_{\rm trans}=50$ (solid), and $\sigma_{\rm trans}=100$ (dashed). The simulation data were time and azimuth-averaged over the final $2500\,t_{\rm g}$, as in \autoref{fig:phiavg}.}
\label{fig:phiavgrsliapp}
\end{figure*}

\section{Simulations with different $\sigma_{\rm trans}$.}
\label{app:sigtranscompare}

In this Appendix, we show results from an additional two MAD simulations run with our new hybrid method, varying the transition magnetization $\sigma_{\rm trans}.$ The simulation parameters and initial conditions are the same as the hybrid simulation in \autoref{sec:simulations}, including the initial random pressure perturbations. In addition to the fiducial hybrid simulation with $\sigma_{\rm trans}=50$ presented in the main text, we ran additional simulations with $\sigma_{\rm trans}=25$ and $\sigma_{\rm trans}=100$. 

Our simulation set-up is stable at all three values of $\sigma_{\rm trans}$. In \autoref{fig:scalarsapp}, we show the accretion rate $\dot{M}$ and MAD parameter $\phi$ from $t=7500\,t_{\rm g}$ to $t=10000\,t_{\rm g}$ for the three simulations. The median $\dot{M}$ and $\phi$ for the $\sigma_{\rm trans}=25$ and $\sigma_{\rm trans}=100$ simulations both differ from the values from the fiducial $\sigma_{\rm trans}=50$ simulation by less than $10\%$.

In \autoref{fig:phiavgrsliapp}, we show radial slices of the averaged magnetization and plasma-$\beta$ in the three hybrid simulations. All three hybrid simulations achieve very high $\sigma$ (low $\beta$) in the jet region close to the black hole. However, the highest value of $\sigma$ (lowest value of $\beta$) achieved in the jet depends on the choice of $\sigma_{\rm trans}$, with  the $\sigma_{\rm trans}=100$ simulation reaching the highest overall $\sigma$ (lowest overall $\beta$) in the jet region. All three simulations have closely matched $\sigma$ and $\beta$ profiles with each other and with the fiducial MHD simulation outside the $\sigma_{\rm trans}$ threshold.

\bsp	
\label{lastpage}

\end{document}